- Review -

# Multifield mediation of the hydrogen-bonding network and properties of the deionized water and the monovalent Lewis-Hofmeister solutions

Chang Q Sun[1,*]

**Abstract**

Phonon spectrometrics examination of the effect of pressure, temperature, confinement (molecular undercoordination), and charge injection by acid, base, and salt solvation (Lewis and Hofmeister solutions) verifies the regulations for the hydrogen bonding (O:H–O or HB with ":" being the lone pair on $O^{2-}$ anion) network and the properties of the deionized water and the solutions. Consistency between theory and measurements confirms the essentiality of the quasisolid phase of negative thermal expansion due to O:H–O segmental specific heat disparity, supersolid phase due to electrostatic polarization by ions injection and molecular undercoordination. Lewis acid and base solvation creates the H↔H anti–HB due to the excessive protons and O:⇔:O super–HB because of the excessive lone pairs, respectively. The multifield meditation of the HB network results in anomalies of water ice and aqueous solutions such as the ice friction, ice floating, superheating and supercooling, warm water speedy cooling, and critical conditions for phase transition. The coupling of inter- and intra-molecular interaction would form important impact to molecular crystals and to the negative thermal expansion of other substance.

[1] EBEAM, Yangtze Normal University, Chongqing 408100, China (ecqsun@qq.com); NOVITAS, Nanyang Technological University, Singapore 639798 (ecqsun@ntu.edu.sg)

**Highlight**

- DPS probes HBs fraction and stiffness transition upon multifield perturbation and charge injection.
- H↔H anti–HB fragilization disrupts the acidic solution network and surface stress.
- O:⇔:O super–HB compression softens the solvent H–O but the basic solute H–O bond stiffens.
- An ion clusters, stretches and polarizes HBs to form the supersolid hydration shells.

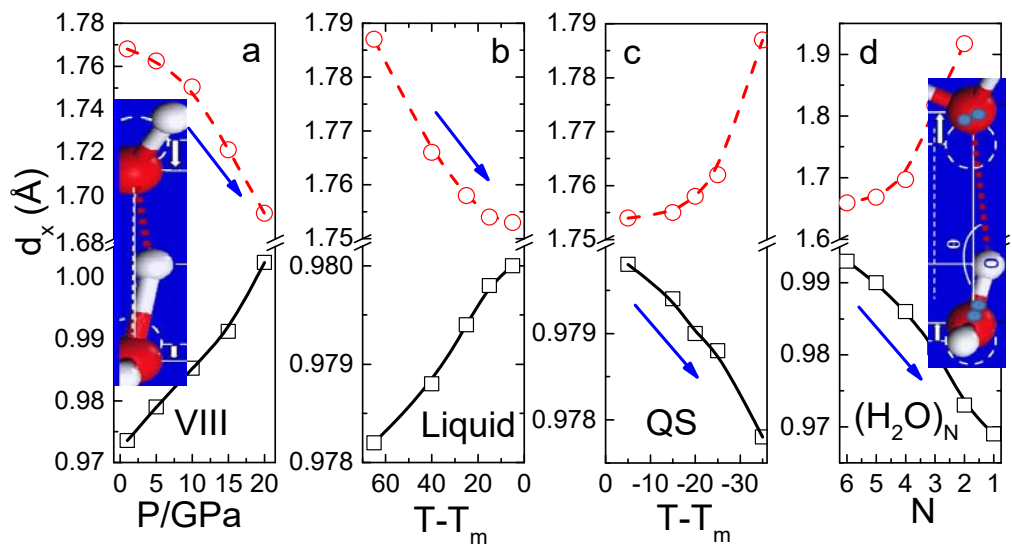

1. Wonders of Water and Aqueous Solution

Water ice responds abnormally to perturbation such as compression, heating, electromagnetic radiation, and molecular undercoordination (or called confinement) leading to anomalies such as ice friction, ice floating, regelation, supercooling/superheating, and warm water cooling faster [1]. The local physical–chemical properties of O:H–O bonds (or HB with ":" being the electron lone pair of oxygen) in the hydration shells of Lewis and Hofmeister solutions are quite different from those of the ordinary bulk water, which has attracted extensive research interest from various perspectives such as molecular dynamics, classical thermal dynamics, and lately O:H–O bond cooperativity [2-5].

Fine–resolution detection and consistently deep insight into the intra– and inter-molecular interactions and their consequence on the hydrogen–bond network and the properties of a solution have been an area of active study. Intensive pump–probe ultrafast photoelectronic and phonon spectroscopic investigations have contributed significantly to pursue the mechanism behind molecular performance in the spatial and temporal domains. For instance, the sum frequency generation (SFG) spectroscopy resolves information on the molecular dipole orientation or the skin dielectrics, at the air–solution interface [6, 7], while the ultrafast two–dimensional infrared absorption probes the solute or water molecular diffusion dynamics in terms of phonon lifetime, population decay, or vibration energy dissipation, and the viscosity of the solutions [8, 9]. The ultrafast photoelectron water–jet spectroscopy reveals the molecular site and droplet size resolved the vertical bound energy and the life time of the hydrated electrons [10, 11]. Theoretical and computations from various perspectives have contributed to the understanding of solvation dynamics, which include $H^+$, $OH^-$, and electron lone pair ":" acceptance and donation, etc. [12-15].

The understanding of solvation can be traced back to 1900's. Svante Arrhenius [12] won the 1903 Nobel prize for his definition of acid-base dissolution in terms of proton $H^+$ and $OH^-$ donation. After some 20 years, Brønsted–Lowry [13, 14] and Lewis [15] defined subsequently the compound dissolution in terms of $H^+$ or electron lone pair ":" donation. Tremendous work has been done since then from various perspectives with derivatives of numerous theories debating mainly on the modes of solute drift motion, hydration shell size, interfacial dielectrics, phonon relaxation time, etc., – given birth of the modern solvation molecular dynamics, or liquid-state science [16]. Charge injection by salt solvation demonstrates the Hofmeister effect [17, 18] on regulating the solution surface stress and the solubility of proteins with possible mechanisms of structural maker and breaker [19-21], ionic specification [22], quantum dispersion [23] fluctuation [24], skin induction [25], and solute–solvent interactions [26]. The performance of the excessive $H^+$ protons and lone pairs in Lewis acid-base solutions has been approached in terms of "molecular structural diffusion" [27] with involvement of proton/lone–pair thermal hopping [28], proton tunneling [29] or fluctuating [30] with polar alteration from proton to lone–pair. Grotthuss [27, 31], Eigen [32], Zundel [33] and their combinations [34-36] explained the possible manner of proton and lone pair transportation, as illustrated in Figure 1.

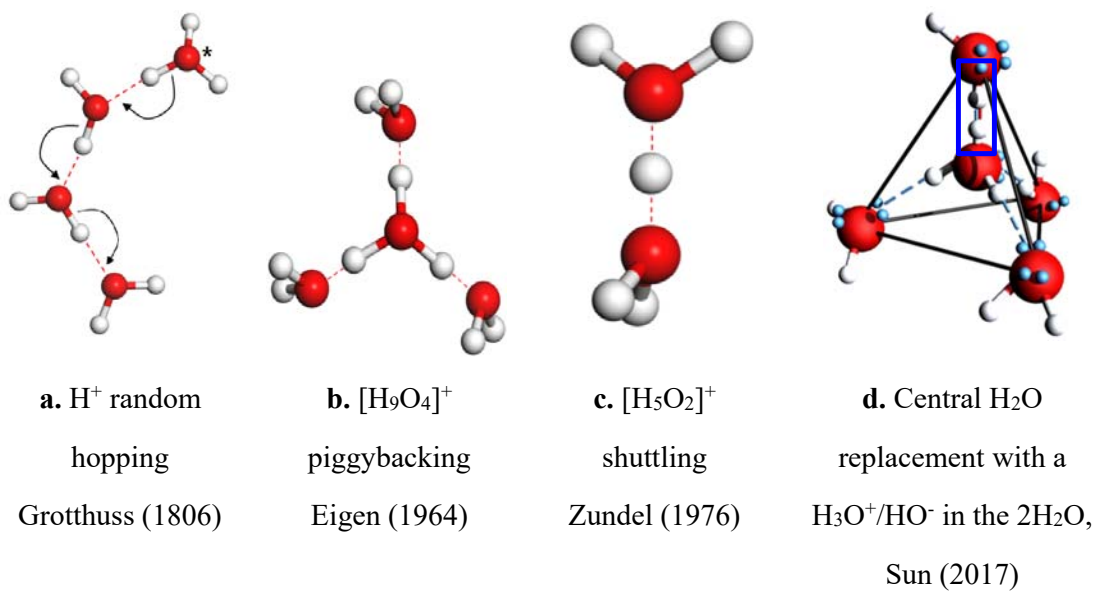

**a.** $H^+$ random hopping
Grotthuss (1806)

**b.** $[H_9O_4]^+$ piggybacking
Eigen (1964)

**c.** $[H_5O_2]^+$ shuttling
Zundel (1976)

**d.** Central $H_2O$ replacement with a $H_3O^+/HO^-$ in the $2H_2O$,
Sun (2017)

Figure 1. Typical models for the proton or lone pair mobility in Lewis solutions by polarity inversion. (a) Grotthuss [27, 31] notion of proton/lone-pair concerted random hopping, (b) Eigen [32] complex $[H_9O_4]^+$ scheme, and (c) Zundel $[H_5O_2]^+$ configuration of proton/lone-pair shuttling (nonbonded [33] or covalently bonded [36] to the neighboring $H_2O$ molecules), and (d) the central $H_2O$ replacement with a $H_3O^+$ for acidic [37] and with a $HO^-$ for the basic solutions [38].

The excessive protons in acidic solutions, and as an inverse of protons, electron lone pairs in basic solutions, form an $H_9O_4^+$ complex in which an $H_3O^+$ core is strongly hydrogen–bonded to three $H_2O$ molecules and leave the lone pair of the $H_3O^+$ free [32], or form an $H_5O_2^+$ complex in which the proton is shuttling freely between two $H_2O$ molecules [33]. The $H_3O^+$ is strongly nonbonded to three $H_2O$ molecules and its lone pair remains free in Eigen's notion. The $H^+$ shuttles between two $H_2O$ molecules in Zundel's scheme. In complementary, this group of practitioners proposed and verified the essentiality of the central $H_2O$ replacement with a $H_3O^+$ or a $HO^-$ to form the H↔H anti-HB point breaker and O:⇔:O super-HB point compressor in Lewis solutions [5, 37, 39]. The $H^+$ or the ":" does not stand alone or freely shuttling between $H_2O$ molecules but forms a $H_3O^+$ or a $HO^-$ tetrahedron firmly. The $H_3O^+$ or $HO^-$ and their short-range surroundings may be subject to Brownian or drift motion under gradient thermal or electric field.

The O:H–O bond cooperativity notion [40] and the differential phonon spectrometrics (DPS) strategy has improved largely the understanding of solvation bonding dynamics, solute capabilities, and inter– and intramolecular interactions in the HX, YX, and YOH solutions (X = Cl, Br, I; Y = Li, Na, K, Rb, Cs) [5, 38, 39, 41-45]. One can resolve the network O:H–O bond segmental cooperative relaxation induced by charge injection in the form of electrons, protons, lone pairs, ions and molecular dipoles upon acid, base, salt solvation, or solute bond–order–deficiency (the bond order of a $HO^-$ with one H–O bond is lower than a $H_2O$ with two H–O bonds). It becomes clear how the $H^+$(or $H_3O^+$), $OH^-$, $Y^+$ and $X^-$ ions interact with water molecules and their neighboring solutes, and their impact on the performance of the solutions such as the surface stress, solution viscosity, solution temperature, and critical pressures and temperatures for phase transition [38, 46]. Compared with focus on the mode of proton/lone-pair or molecular drifting, what is going on inside the water molecules is much more fascinating [37, 38, 42].

Aside from the interest in the O:H phonon relaxation [47], solute–solvent interaction length [48], structure making or breaking [49, 50], solute motion dynamics and phonon relaxation lifetime, one needs to examine the intra– and intermolecular interactions and the solute capabilities of transiting the number and stiffness of the O:H–O bonds from the mode of ordinary water into the hydration shells. The DPS strategy and the HB transition theory enabled verification of that O:H–O bond polarization, H↔H interproton disruption, and O:⇔:O inter–lone–pair compression essentially govern the solute–solvent interactions in the respective H(Cl, Br, I)[37], Na(F, Cl, Br, I)[43] and (LI, Na, K)OH[42] solutions.

The aim of this presentation is to show how the multifield mediate the HB network and the properties of water and Lewis-Hofmeister solutions with the aid of the DPS strategy. The DPS resolves the O:H–O bond transition from the mode of ordinary water to its hydration in terms of its phonon stiffness (vibration frequency shift $\Delta\omega$), order of fluctuation (line width), and number fraction (phonon abundance, $f(C) = N_{hydration}/N_{total}$ is the DPS peak area integration) of HB polarization. Therefore, one can correlated the macroscopic properties of a solution to the fraction of bond transition upon charge injection or multifield perturbation.

2. O:H−O Bond Cooperativity and Segmental Specific Heat Disparity

2.1. Basic Rules for Water

Water prefers the statistic mean of the tetrahedrally–coordinated, two–phase structure in a core–shell fashion of the same geometry but different O:H−O bond lengths [1, 51]. Figure 2a illustrates the 2H$_2$O unit cell of C$_{3v}$ symmetry having four hydrogen bonds bridging oxygen anions. It is essential to treat water as a crystalline–like structure with well–defined lattice geometry, strong correlation, and high fluctuation. For a specimen containing N number of O$^{2-}$ anions, there are 2N protons H$^+$ and 2N lone pairs ":" and the O:H−O bond configuration conserve regardless of structural phase [52] unless excessive H$^+$ or ":"is introduced [37, 39].

The motion of a H$_2$O molecule or the proton H$^+$ is subject to restriction. If a molecule rotating 60° and above around its C$_{3v}$ symmetrical axis, there will be H↔H anti–HB and O:⇔:O super–HB formation that is energetically forbidden. Because of the H−O bond energy of ~4.0 eV, translational tunneling of the H$^+$ is also forbidden. Breaking the H−O or the D−O bond in vapor phase requires 121.6 nm laser radiation [53, 54], estimated 5.1 eV because the extremely low molecular coordination numbers.

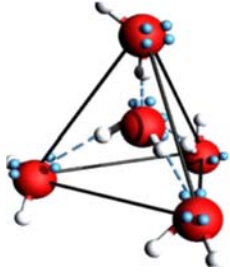

(a) 2H₂O unit cell     (b) O:H−O specific heat     (c) ρ(T) oscillation

Figure 2. (a) The 2H$_2$O primary unit cell contains four oriented O:H−O bonds defines liquid water as a crystal with molecular and proton motion restrictions [51]. (b) Superposition of the segmental specific heat $\eta_x$ defines (c) the phases of density oscillation, for bulk water (T ≥ 273 K) and 1.4 nm water droplet (T < 273 K) [55], from high temperature downward of the Vapor ($\eta_L$ = 0, not shown), Liquid and ice I$_{h+c}$ ($\eta_L/\eta_H$ < 1), quasisolid (QS) ($\eta_L/\eta_H$ > 1), XI ($\eta_L \cong \eta_H \cong$ 0), and the QS boundaries ($\eta_L/\eta_H$ = 1) closing to T$_m$ and T$_N$, respectively [4]. Electrification (ionic polarization)[5] or molecular undercoordination [3] disperses the QS boundaries outwardly but compression inwardly [56] governed by Einstein relationship: $\Theta_{Dx} \propto \omega_x$ (x = L, H for the O:H and the H–O bond). Negative thermal expansion occurs only to the QS phase [4].

2.2. Specific Heat and Density Oscillation

It is necessary to decompose the specific heat of water ice into two components $\eta_x(T/\Theta_{Dx})$ of Debye approximation (X = L and H for the O:H and H–O segment, respectively) [4]. The segmental specific heat meets two criteria. One is the Einstein relation $\omega_x \propto \Theta_{Dx}$ and the other is its thermal integration being proportional to bond energy $E_x$. The ($\omega_x$, $E_x$) is (200 cm$^{-1}$, ~0.1 eV) for the O:H nonbond and (3200 cm$^{-1}$, ~4.0 eV) for the H–O bond. The Debye temperatures and the specific heat curves are subject to the $\omega_x$ that varies with external perturbation. Figure 2b shows the superposition of the specific heats $\eta_x$, which defines the phases in Figure 2c showing density oscillation over the full temperature regime [4].

The hydrogen bonding thermodynamics at a certain temperature is subject to the specific heat ratio, $\eta_L/\eta_H$. The segmental having a lower specific heat follows the regular thermal expansion but the other segment responds to thermal excitation oppositely because of the HB cooperativity. As illustrated in Figure 3 insets. The O ions dislocate in the same direction but by different amounts.

In the Vapor phase, $\eta_L \cong 0$, the O:H interaction is negligible; in the Liquid and $I_{c+h}$ ice, $\eta_L/\eta_H < 1$, O:H cooling contraction takes place at different rate, but the H–O elongates less than the O:H does. In the XI phase, $\eta_L \cong \eta_H \cong 0$, neither O:H nor H–O responds sensitively to temperature change. At the QS boundaries ($\eta_L/\eta_H = 1$) density transits from elongation/contraction to contraction/elongation. In the QS phase, $\eta_L/\eta_H > 1$, H–O cooling contraction and O:H expansion take place; which triggers ice floating when cooling at the QS phase because the H–O contracts less than O:H expansion at cooling [4].

2.3. Supersolidity and quasisolidity

The concept of supersolidity was initially extended from the $^4$He fragment at mK temperatures, demonstrating elastic, repulsive and frictionless between the contact motion of $^4$He segments [57] because of atomic undercoordination induced local densification of charge and energy and the associated polarization [58]. The concepts of supersolidity and quasisolidity were firstly defined for water and ice in 2013 by this group [3, 4] and then intensively verified subsequently.

The quasisolidity describes phase transition from Liquid density maximum of one gcm$^{-3}$ at 4 °C to Solid density minimum of 0.92 gcm$^{-3}$ at –15 °C, which demonstrates the cooling expansion because the specific heat ratio $\eta_L/\eta_H < 1$, the H–O bond contraction drives the O:H expansion and the ∠O:H–O angle relaxation from 160 to 165°. The QS boundaries are subject to dispersion by external stimulus, which is why water ice perform anonymously in thermodynamics.

The supersolidity features the behavior of water and ice under polarization by coordination number (CN or z) reduction or electric polarization [3, 5]. When the nearest CN number is less than four the H–O bond contracts spontaneously associated with O:H elongation and strong polarization. At the surface, the H–O bond contracts from 1.00 to 0.95 Å and the O:H expands from 1.70 to 1.95 Å associated with the O:H vibration frequency transiting from 200 to 75 cm$^{-1}$ and the H–O from 3200 to 3450 cm$^{-1}$ [59]. Salt solvation derives cations and anions dispersed in the solution [40]. Each of the ions serves as a source center of electric field that aligns, stretches and polarizes the O:H–O, resulting the same supersolidity in the hydration shell whose size is subject to the screening of the hydrating $H_2O$ dipoles and the ionic charge quantity and volume size.

The shortened H–O bond shifts its vibration frequency to a higher value that increases further with the reduction of the molecular CN, which disperses the QS boundaries outwardly, causing the supercooling at freezing and superheating at melting [15]. Under compression, the situation reverses, raising the $T_N$ and lowering $T_m$, which is the case of regelation – ice melts under compression and the $T_m$ reverse when the pressure is relieved [49]. The high thermal diffusivity of the supersolidity skin governs thermal transportation in the Mpemba paradox – warm water cools faster.

2.4. O:H–O Length-stiffness-energy Relaxation

2.4.1.   Segmental Length

As the basic structure and energy exchange unit, the O:H−O bond integrates the intermolecular weaker O:H nonbond (or called van der Waals bond with ~0.1 eV energy) and the intramolecular stronger H−O polar–covalent bond (~4.0 eV), rather than either of the O:H or the H–O alone, with asymmetrical and short–range interactions and coupled with the Coulomb repulsion between electron pairs on adjacent oxygen ions [51]. O:H−O length and bond angle relaxation changes system energy, but fluctuation contributes little on an average.

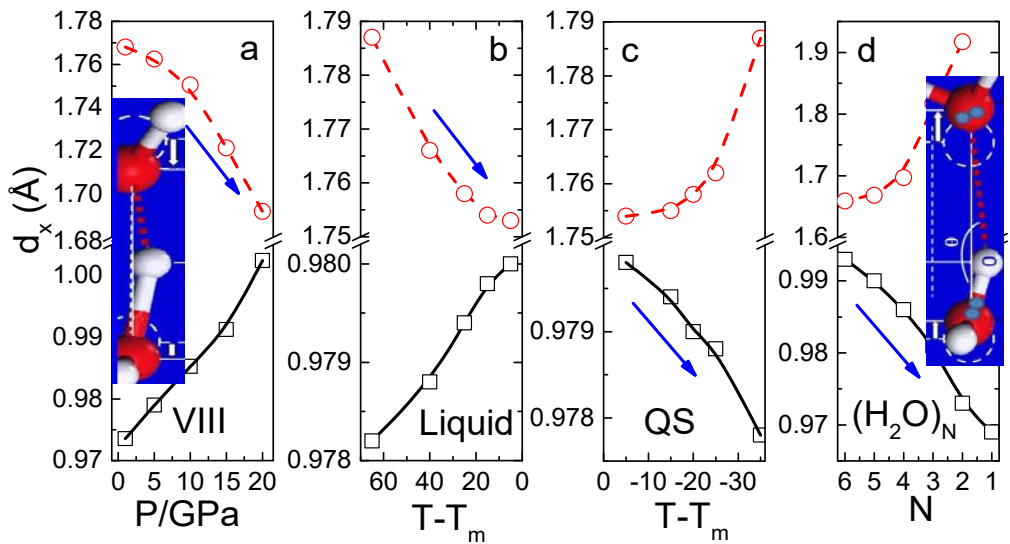

Figure 3. MD-derived O:H–O cooperative relaxation under (a) mechanical compression, (b) Liquid cooling, (c) QS cooling, and (d) $(H_2O)_N$ cluster molecular undercoordination. Arrows denote the master segments and their relaxation directions. The O:H always relaxes more than the H–O in opposite slopes and curvatures, irrespective of the stimulus applied or the structural phase because of the persistence of O–O repulsive coupling [51]. Calculations were performed using the force field of Sun [60].

The O:H nonbond and the H−O bond segmental disparity and the O−O coupling allow the segmented O:H−O bond to relax oppositely – an external stimulus dislocates both O ions in the same direction but by different amounts, see Figure 3. The softer O:H nonbond always relaxes more than the stiffer H−O bond with respect to the $H^+$ coordination origin. The ∠O:H−O angle θ relaxation contributes only to the geometry and mass density. The O:H−O bond bending has its specific vibration mode that does not interfere with the H−O and the O:H stretching vibrations [51]. The O:H−O bond cooperativity determines the properties of water and ice under external stimulus such as molecular undercoordination [61-65], mechanical compression [2, 38, 46, 66, 67], thermal excitation [4, 68, 69], solvation [70, 71] and determines the molecular behavior such as solute and water molecular thermal fluctuation, solute drift motion dynamics, or phonon relaxation.

2.4.2. Segmental Phonons

Figure 4 shows the $\omega_x$ DPS for deionized water subjected to (a, b) heating [72], (c) compression [38] and (d) molecular undercoordination [59, 73]. Heating stiffens the $\omega_H$ from 3200 to 3500 cm$^{-1}$ and meanwhile softens the $\omega_L$ from 180 to 75 cm$^{-1}$. The O:H nonbond in liquid water follows the regular rule of thermal expansion but the H−O bond is subjecting to thermal contraction because of the O−O repulsive correlation [4]. Mechanical compression softens the $\omega_H$ from the skin value of 3450 to 3100 cm$^{-1}$, which confirms that mechanical compression shortens the O:H nonbond and lengthens the H−O bond. The compression effect retains regardless of the structure phase of water [38] or ice [2]. These observations prove the O−O repulsivity that correlates the inter- and intra-molecular interactions.

The $\omega_H$ DPS of water and ice gained by varying the angle between the surface normal and the direction of light reflection distils the monolayer-skin bonding information of (d) 25 °C water and ice (-20 and -15 °C). The phonon transits its frequency from the bulk water (3200 cm$^{-1}$) and bulk ice (3150 cm$^{-1}$) to their skins sharing the identical H−O bond stiffness featured at 3450 cm$^{-1}$, which clarifies that the length and energy of the H−O bond in both skins are identical disregarding temperature or structure phase [59]. An integral of the skin DPS abundance suggests that the skin of ice is 9/4 times thick of liquid water. However, heating reduces the surface stress and the contact angle between water and glass shown inset **a** as a result of thermal fluctuation [74]. Table 1 compares the O:H-O segmental length and stiffness relaxation under perturbation.

Strikingly, measurements in panel (d) confirm the "molecular undercoordination resolved two-phase structure of water and ice" [1]. Covered with a supersolid skin, water is a uniform, strongly correlated, fluctuating crystal-like. This observation is consistent with the wide- and small-angle X-ray scattering observations from amorphous ice [75] and liquid droplet [76]. Nanodroplets and amorphous states are naturally the same as they share undercoordinated molecules in different manners. The only difference between amorphous and a nanostructure is the distribution of the undercoordinated defects - one is ordered at the domain skins and the other is randomly distributed in the bulk. Therefore, the two-phase structure, low-density supersolid skin and the normal bulk, holds for water and ice. This observation may clarify the long-debating two-phase structural model as "coordination-resolved" other than "domain-resolved" nature [75, 76]. A spectroscopy resolves the atomic distance or bond vibration frequency disregarding their locations or orientations in the real space.

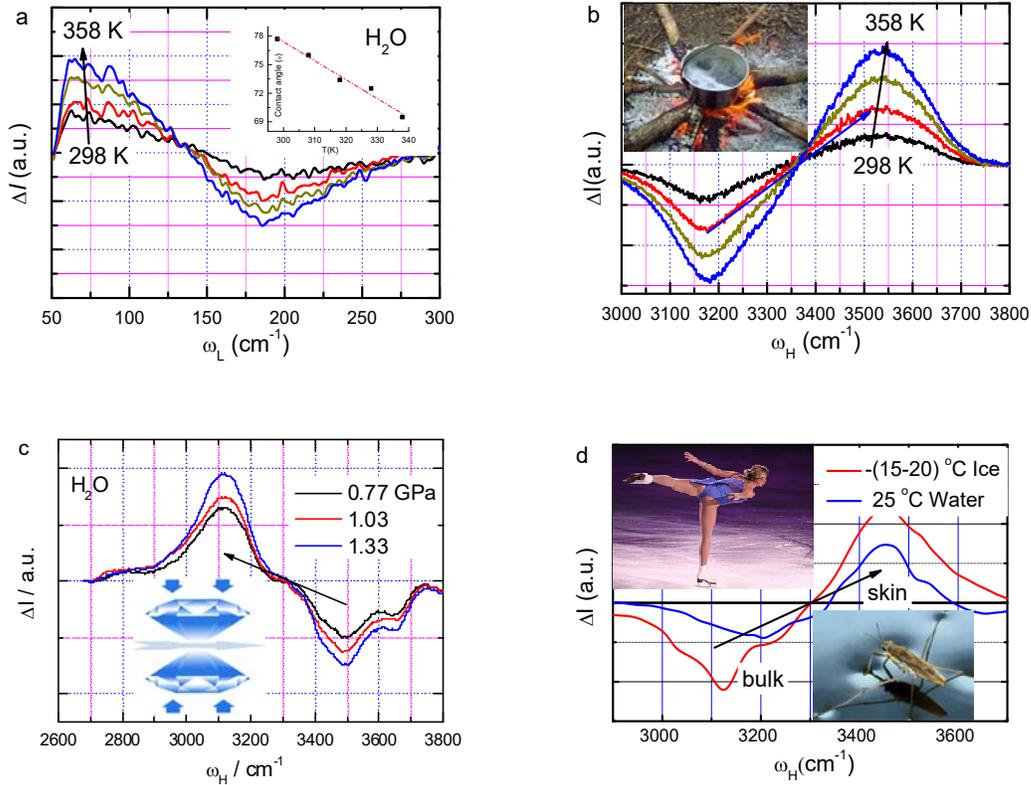

Figure 4. Thermal excitation (a) softens the $\omega_L$ from 180 to 75 cm$^{-1}$ and (b) stiffens the $\omega_H$ from 3200 to 3500 cm$^{-1}$ for deionized water [72]. (c) Mechanical compression transits the $\omega_H$ of liquid water from above 3300 cm$^{-1}$ to its below [38] and, (d) molecular undercoordination transits the $\omega_H$ from 3200 for water (at 25 °C) and 3150 cm$^{-1}$ for ice (-20 and -15 °C) to 3450 cm$^{-1}$ for the skins of water and ice [59, 73]. Inset a shows the thermal depression of the contact angle that is the same in trend of surface tension [74].

Numerical reproduction of the Mpemba effect – hot water cools faster [72], evidences directly the essentiality of the 0.75 unit mass density of the supersolid skin that promotes heat conduction outward the water of heat source. Exothermic reaction proceeds by bond elongation and dissociation while endothermic reaction proceeds by bond contraction and bond formation. The Mpemba effect integrates the O:H−O bond energy "storage-emission-conduction-dissipation" cycling dynamics. The energy storage is proportional to the H−O bond heating contraction and the rate of energy emission at cooling is proportional to its first storage. The skin higher thermal conductivity due to lower mass density benefits heat flow outward the solution, and the source-drain non-adiabatic dissipation ensures heat loss at cooling.

Table 1. O:H–O segmental cooperative relaxation in length, vibration frequency, and surface stress with respect to $d_{L0}$ = 1.6946 Å, $d_{H0}$ = 1.0004 Å, $\omega_{H0}$ = 3200 cm$^{-1}$, $\omega_{L0}$ = 200 cm$^{-1}$, $\Theta_{DH}$ = 3200 K, $\Theta_{DL}$ = 198 K upon excitation by heating, compression, molecular undercoordination (skin, cluster, droplet, nanobubble). ($\Delta\Theta_{Dx} \propto \Delta\omega_x$)

| | $\Delta d_H$ | $\Delta d_L$ | $\Delta\omega_H$ | $\Delta\omega_L$ | Remark | Ref |
|---|---|---|---|---|---|---|
| $\Delta T > 0$ (277, 377 K) | <0 | >0 | >0 | <0 | Liquid and solid thermal expansion | [4] |
| $I_c + I_h$ (100, 258 K) | | | | | | |
| $\Delta T < 0$ (258, 277 K) | >0 | <0 | <0 | >0 | QS negative thermal expansion | |
| XI (0, 100 K) | ≅0 | | | | Debye specific heat ≅ 0 | |
| $\Delta z < 0$; $\Delta E \neq 0$ (charge) | <0 | >0 | >0 | <0 | polarization; supersolidity | [5] [3] |
| $\Delta P > 0$ | >0 | <0 | <0 | >0 | $d_L$ and $d_H$ symmetrization | [38] |

2.4.3. Potential Paths of the Relaxing O:H–O bond

Figure 5a illustrates the asymmetrical, short–range, coupled three–body potentials for the segmented O:H−O bond [77, 78]. The proton serves as the coordination origin. The left–hand side is the O:H van der Walls (vdW) interaction and the right–hand side is the H−O polar covalent bond. The Coulomb repulsion between electron pairs on neighboring O$^{2-}$ couple the O:H−O bond to be an oscillator pair.

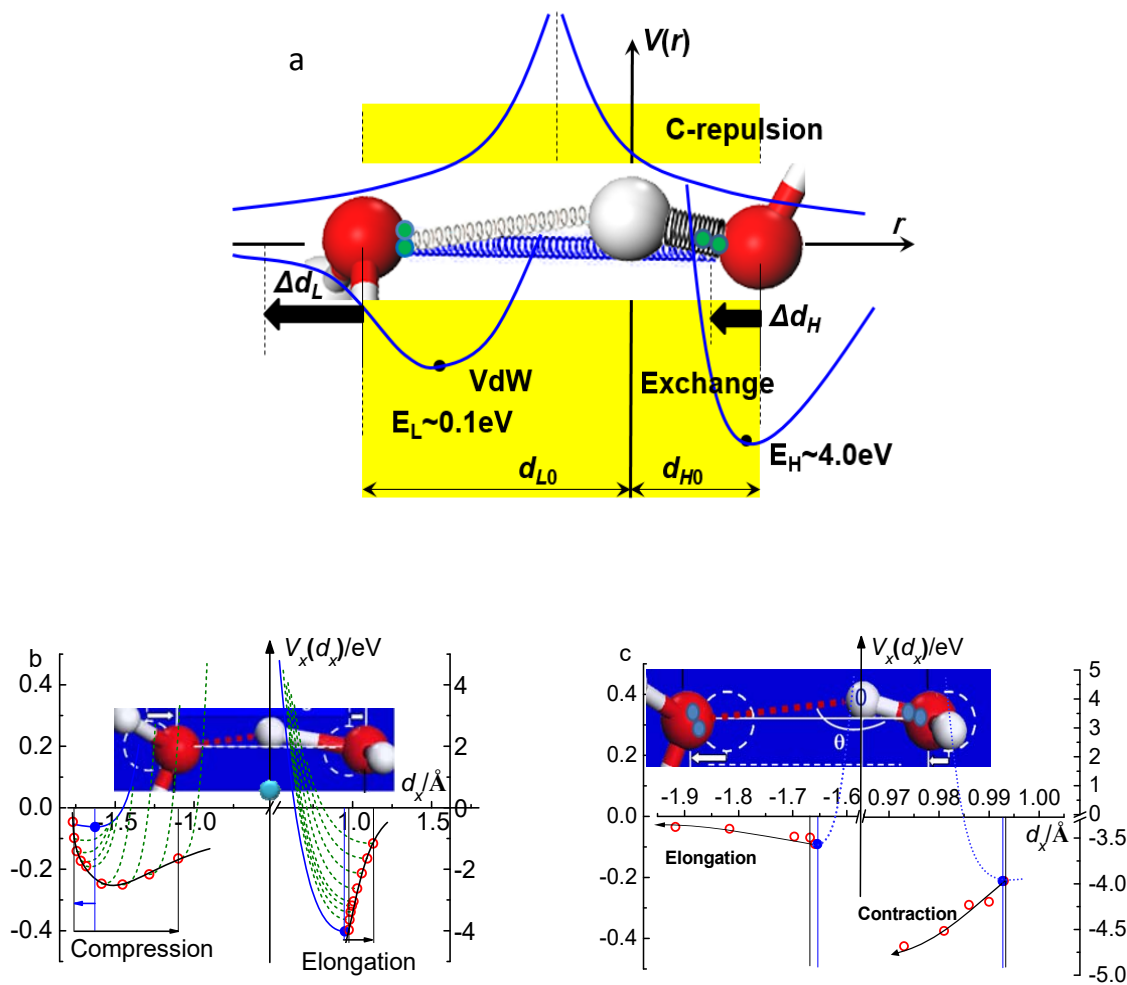

Figure 5. (a) The asymmetrical, short–range, coupled three–body potentials for the segmented O:H−O bond oscillator pair [77, 78]. Potential paths (red circles) for the (b) compression shortened [77] and (c) molecular undercoordination elongated [78] O:H−O bond, derived from Lagrangian-Laplacian transition from the measured segmental lengths and vibration frequencies to the force constants and bond energies of the coupled oscillator under the external perturbations (Reprinted with permission from [77, 78].)

Figure 5b and c is the Lagrangian-Laplacian transition from the measured segmental lengths and vibration frequencies ($d_x$, $\omega_x$) at each point of equilibrium to the force constants and bond energies of the coupled oscillator under (b) mechanical compression of 80 K ice (r. to l.: P = 0 to 60 at 5 GPa step)[77] and (c) undercoordination of $(H_2O)_N$ clusters at the ambient (r. to l.: N = 6, 5, 4, 3, 2). Blue dots are states at the V′x = 0 equilibrium without involvement of O−O repulsion. Red circles in the leftmost (a) and in the rightmost (b) are states at V′x + V′C = 0 equilibrium with the O−O repulsion being involved. Rest red circles are states subjecting to the V′x + V′C + f$_{ex}$ = 0. The f$_{ex}$ is the non-conservative force due stimulus, V′x is the gradient of the inter- and the intra-molecular potential and V′C is the gradient of O−O Coulomb potential. Both O dislocate in the same direction but by different amounts. The O:H−O relaxation proceeds only in one of the two manners under any stimulation [51]:

1) HB segmental disparity and O−O Coulomb repulsivity define the segmental potentials and their extraordinary adaptivity, cooperativity, recoverability, and sensitivity when responding to perturbation.
2) HB elongation occurs under electrification (ionic polarization of the O:H−O bond), hydrophobic capillary confinement, molecular undercoordination, liquid and solid heating, quasisolid cooling, or mechanical tension, which reduces the H$_2$O molecular size (d$_H$) but enlarges their separations (d$_L$) associated with stiffening the H−O bond and softening the O:H oscillator.
3) HB contraction in the opposite manner takes place under mechanical compression, base solvation, liquid and solid cooling, quasisolid heating, which enlarges the d$_H$ but reduces the d$_L$ associated with softening of the H−O bond and stiffening the O:H oscillator.
4) The unprecedented O:H−O bond cooperative relaxation arise from the O−O coupling and the externally applied non-conservative activation.

3. Lewis and Hofmeister Solutions

3.1. Solute–solvent Nonbonding Interaction

Figure 6 insets illustrate the charge injection by (a) salt, (b) acid, and (c) base solvation. Charge injection by solvation mediates the hydrogen bonding network and the properties of a Hofmeister solution through the electrostatic polarization and hydration shell formation with H$_2$O shielding in the hydration volume. H↔H anti-HB fragilization, O:⇔:O super-HB compression and solute H−O bond contraction present to the Lewis acid and basic solutions.

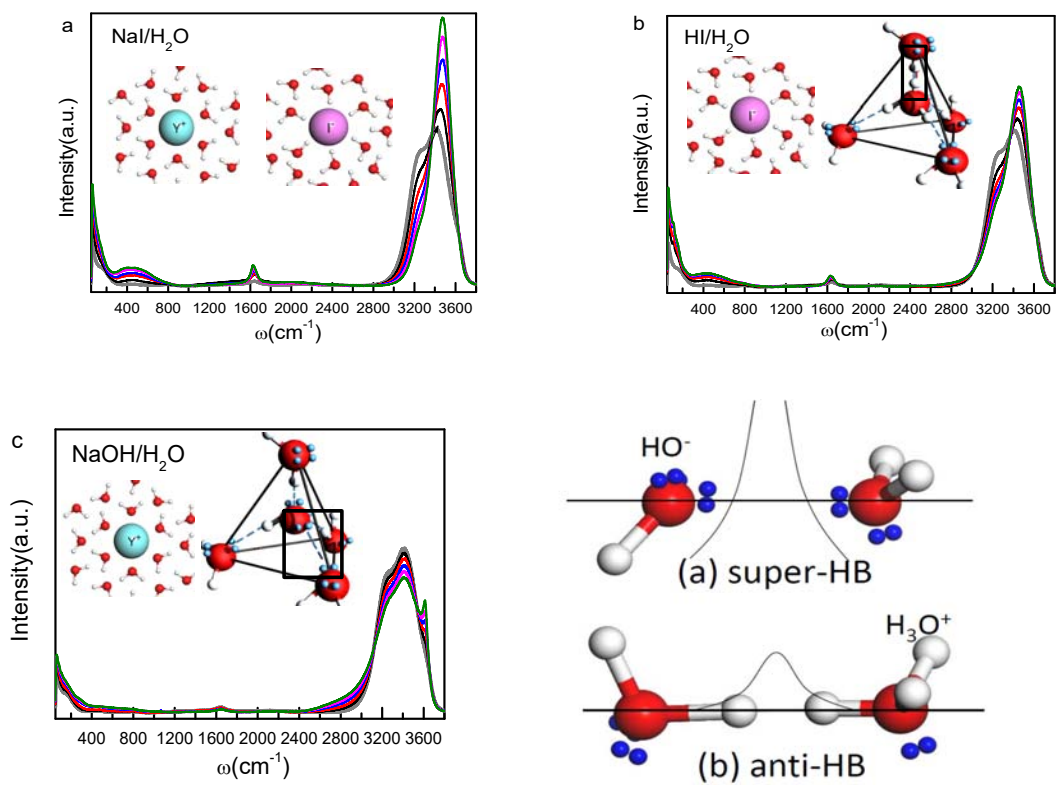

Figure 6. Full–frequency Raman spectra for the concentrated (a) NaI/$H_2O$ [38], (b) HI/$H_2O$ [37], and (c) NaOH/$H_2O$ [42] solutions. (d) Illustration of the O:⇔:O super–HB between the $OH^-$ and a $H_2O$ and the H↔H anti–HB between the $H_3O^+$ and a $H_2O$ molecule and their respective repulsive potentials. Inset a illustrates the effect of charge injection by salt solvation. Insets b and c show the central $H_2O$ replacement by $H_3O^+$ in acid and $HO^-$ in basic solutions. Using molar fraction C = $N_{solute}/N_{solvent}$ will be more convenient than using the molarity mol/L convention and C = (mol/L)/(1000/18 + mol/L) (Reprinted with copyright permission from [37, 38, 44, 51, 79]).

Salt solvation derives cations and anions dispersed in the solution, except for the closed ion pairs that performs like a dipole [5]. Each of the ions serves as a source center of electric field that aligns, stretches and polarizes the O:H–O bond, resulting in the supersolid hydration shell whose size is subject to the screening of the hydrating $H_2O$ dipoles and the ionic charge quantity and volume size. $X^-$ and $Y^+$ ionic injection does not break the 2N number conservation of pure water, but each ion serves as a point polarizer that was surrounded by a limited number of $H_2O$ dipoles in the supersolid [4] or semirigid [49, 80] hydration shells. Ion injection only distorts the local structure with covalent bond being formed between solute and solvent, as Figure 6a inset illustrated [43].

However, an introduction of an excessive ":" or $H^+$ braeks the 2N number and the O:H–O configuration invariance. For instance, LiOH solvation adds a $HO^-$ with one $H^+$ and three ":", turning the 2N protons into 2N+1 and the 2N lone pairs into 2N+3, resulting in the 2N+3–(2N+1) = 2 excessive lone pairs that can only form the O:⇔:O interaction without any other choice [42]. Likewise, HBr solvation creates the H↔H interaction [37]. The unprecedtly H↔H interproton repulsion and O:⇔:O inter–lone–pair compression govern the performance of the acid and basic solutions [37, 42].

Both the $H_3O^+$ and the $OH^-$ retain their $sp^3$–hybridized electron orbitals but have unbalanced numbers of protons and lone pairs, as the Figure 6 b and c insets illustrated [37, 42]. The $H_3O^+$ and $OH^-$ substitution for the central $H_2O$ molecule in the $2H_2O$ unit cell creates regularly the (Figure 6b inset) H↔H anti–HB point breaker and the O:⇔:O super–HB point compressor. Figure 6d further illustrates the H↔H anti–HB and O:⇔:O interactions. No such conclusion could be possible if one assumed water as an amorphous substance or a randomly ordered system or considered the $H^+$ or the ":" freely hopping or shuttling. Protons and lone-pair could not stay alone in the liquid but bond to a $H_2O$ molecule to form a $H_3O^+$ and $HO^-$ that may subject to Brownian motion and drift motion under a thermal or electric field gradient.

These point H↔H breakers, O:⇔:O compressors, and ionic polarizers govern the performance of the hydration network of Lewis-Hofmeister solutions. The H↔H anti–HB disrupts the solution network and the surface stress [37], which is the same to the H–induced embrittlement of metals and alloys [81, 82]. The O:⇔:O super–HB compresses the neighboring O:H–O bond [42] to have the same effect of mechanical compression that shortens the O:H nonbond and elongates the H–O bond [38].

3.2. DPS Derived Fraction Coefficient

3.2.1.    (Li, H)Br and LiOH DPS Profiles

The full–frequency Raman spectra, shown in Figure 6, for the concentrated NaI [38], HI [37], and NaOH [42] solutions were collected under the ambient conditions [83]. The Raman spectrum covers the phonon bands of O:H stretching vibration at < 200 cm$^{-1}$, the ∠O:H–O bending band centered at 400 cm$^{-1}$, the ∠H–O–H bending band at 1600 cm$^{-1}$, and the H–O stretching band centered at 3200 cm$^{-1}$. The O:H stretching, molecular rotational and torsional vibrations are within the THz regime, one can hardly discriminate these contributions one from the other. Inset (c) decomposes the H–O stretching phonon band into the bulk (3200 cm$^{-1}$), the skin or the surface having a certain thickness (3450 cm$^{-1}$), and the surface dangling H–O bond or called free radical (3610 cm$^{-1}$) directing outwardly of the surface [51]. Likewise, the O:H stretching vibration phonon centered at 75 cm$^{-1}$ features the undercoordination–induced skin O:H elongation and polarization, the ~200 cm$^{-1}$ peak feature the O:H vibration for the four–coordinated molecules in the ordinary bulk water. From the full–frequency spectra, one can hardly tell the effect of charge injection that stiffens the H–O phonon and softens the O:H and the excessive feature at 3610 cm$^{-1}$ for HO$^-$ and H$_2$O$_2$.

The Raman frequency shift $\Delta\omega_x$, features the stiffness of the segmental x stretching vibration as a function of its length $d_x$ and energy $E_x$ [51], $\Delta\omega_x \propto \sqrt{E_x/\mu_x}/d_x \propto \sqrt{(k_x+k_c)/\mu_x}$. The subscript x = L denotes the O:H nonbond characterized by the stretching vibration frequency at ~200 cm$^{-1}$ and x = H denotes the H–O bond phonon frequency of ~3200 cm$^{-1}$ in the bulk water. The $k_x$ and $k_C$ are the force constants or the second differentials of the intra/inter molecular interaction and O–O Coulomb coupling potentials. The $\Delta\omega_x$ also varies with the reduced mass $\mu_x$ of the specific x oscillator. However, from the full–frequency Raman spectra, one could hardly be able to resolve the transition of bonds by solvation. Inclusion of high–order nonlinear interactions only offsets the peak position without adding any new features of vibrations [51, 84].

The DPS distils only phonons transiting into their hydration states as a peak above the x–axis, which equals the abundance loss of the ordinary HBs as a valley below the axis in the DPS spectrum. This process removes the spectral areas commonly–shared by the ordinary water and the high–order hydration shells. The DPS [85, 86] resolves the transition of the phonon stiffness (frequency shift) and abundance (peak area) by solvation. The fraction coefficient, $f_x(C)$, being the integral of the DPS peak, is the fraction of bonds, or the number of phonons transiting from water to the hydration states at a solute concentration C.

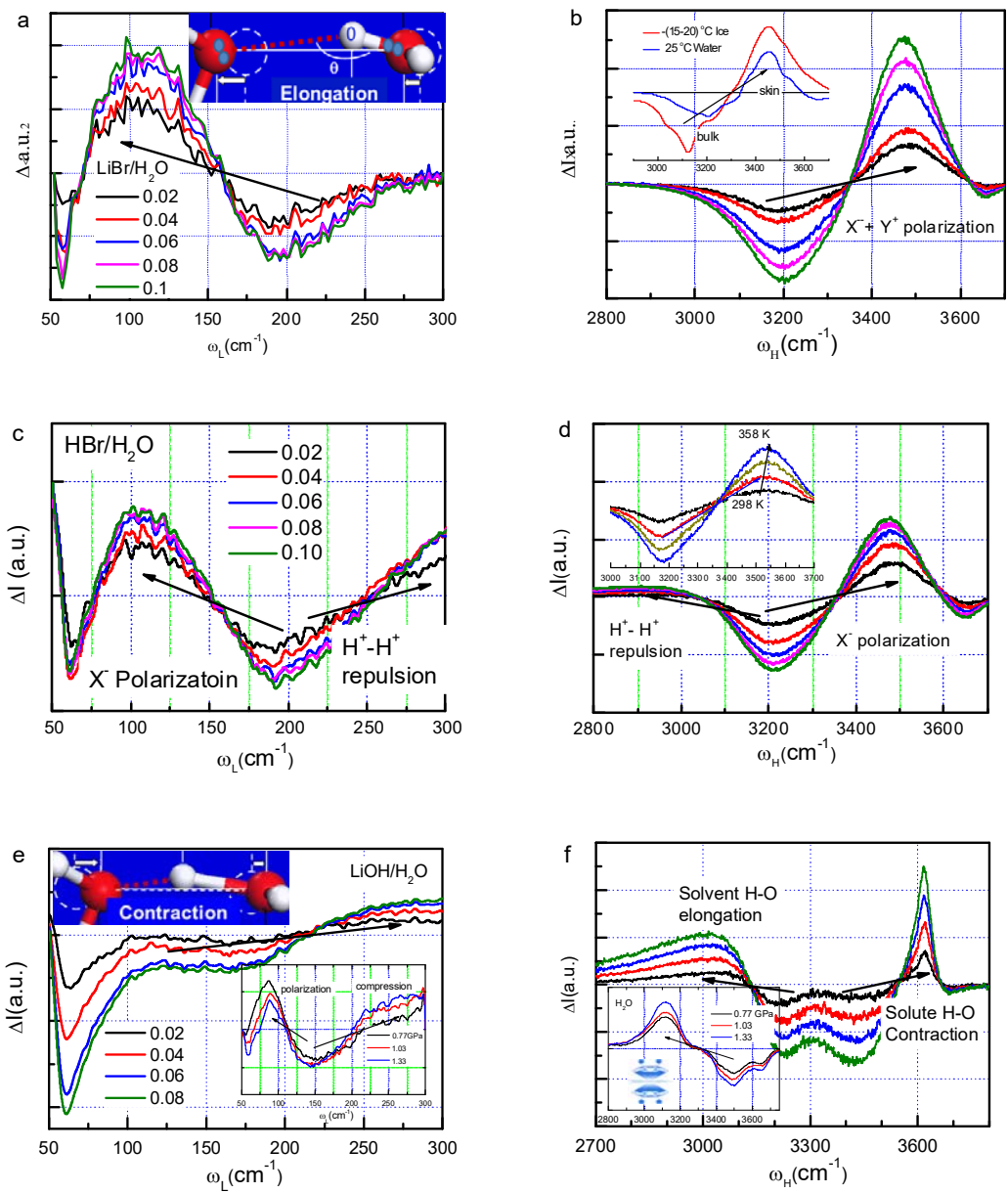

Figure 7. DPS $\omega_x$ for concentrated (a, b) LiBr/$H_2O$ [5], (c, d) HBr/$H_2O$ [37], and (e, f) LiOH/$H_2O$ [42] solutions. Insets a and e illustrate the manners of O:H–O bond elongation by ionic polarization and contraction by O:⇔:O compression, respectively. Inset b shows effect of skin molecular undercoordination and d the effect of liquid water heating. Insets e and f shows the DPS resulting from mechanical compression at room temperature of liquid water. The humps below 3100 cm$^{-1}$ arise from the H↔H repulsion and O:⇔:O compression that elongates the H–O bond of the neighboring water.

### 3.2.2. Effect of Aqueous Charge Injection

From the DPS profiles in Figure 7 for the LiBr, HBr [37] and LiOH [42], one can obtain the fraction coefficients by the peak area integration. The slope of the fraction coefficient, $df_x(C)/dC$, is in proportional to the number of bonds per solute in the hydration shells, which characterizes the hydration shell size and its local electric field. A hydration shell may have one, two or more subshells, depending on the nature and size of the solute. The size and charge quantity determine its local electric field intensity that is subject to the screening by the local $H_2O$ dipoles and modified by the solute–solute interactions [87].

Ionic polarization transits cooperatively the O:H–O segmental phonons $\omega_H$ from 3200 to 3480 cm$^{-1}$ and $\omega_L$ from 200 to 100 cm$^{-1}$ in their hydration shells because of the O–O repulsion, as Figure 7 b inset illustrated. Ionic polarization has the same effect of molecular undercoordination to polarize and transit the O:H–O bond length and stiffness [51, 88]. Therefore, the ionic hydration shells behave identically to the supersolid water skin. The supersolid means highly ordered structure (longer $\omega_H$ lifetime [9]) of semirigid [49], high stress, polarized charge distribution, low density, slow molecular dynamics, and high thermal stability[43, 51].

The difference of the $\omega_H$ phonon abundance between the LiBr and the HBr solution in Figure 7b and d shows that the H–O phonon abundance of the HBr solution is less than the LiBr, which confirmed that the H$^+$ in the HBr solution does not polarize its neighboring $H_2O$ molecules because of the $H_3O^+$ formation. On the other hand, the H↔H repulsion shifts a tiny fraction, but O:⇔:O compression shifts considerable amount of the solvent H–O feature to 3100 cm$^{-1}$ and below. This high broadness indicates the distant solvent O:H–O bond relaxing by the point compression. The <3100 cm$^{-1}$ phonon abundance difference between Figure 7d and f discriminates the strength of the H↔H and the O:⇔:O repulsive interactions. The latter is estimated four time of the former by considering the charge quantities of the same separation.

The DPS for LiOH solution in Figure 7f also shows an excessive sharp peak at 3610 cm$^{-1}$, which indicates the rather local nature of the solute H–O bond contraction. The spectral shift annihilates the effect of Li$^+$ polarization. Excitingly, the O:⇔:O compression is much greater than the critical pressure, 1.33 GPa, for room–temperature water–ice transition. As shown in Figure 7 e and f insets mechanical compression transits the H–O phonon from 3300 cm$^{-1}$ to below [38]. Figure 7 insets compare the effect of (b) molecular undercoordination that transits the $\omega_H$ from 3200 for water (at 25 °C) and 3150 cm$^{-1}$ for ice (–20 and –15 °C) to 3450 cm$^{-1}$ for the skins of water and ice [59, 73], (d) thermal excitation stiffens the $\omega_H$ from 3200 to 3500 cm$^{-1}$ for deionized water [72].

The two DPS peaks clarify that the longer 200 ± 50 fs lifetime features the slower molecular motion but higher–frequency 3610 cm$^{-1}$ solute H–O bond vibration and the other shorter time on 1–2 ps scales is related to the lower–frequency <3100 cm$^{-1}$ elongated solvent H–O bond vibration upon HO$^-$ solvation[42] in NaOH solutions [89, 90].

3.2.3. Solute–Solvent and Solute–Solute Interactions

Figure 8 compares the concentration dependent $f_{LiBr}(C) = f_{Li}(C) + f_{Br}(C)$, $f_{HBr}(C) = f_{Br}(C)$ and $f_{LiOH}$(<3100 cm$^{-1}$, 3610 cm$^{-1}$) that feature the relative number of O:H–O bonds transiting from the ordinary water into the hydrating states. The $f_x(C)$ concentration trends recommend the following, see Figure 8:

1) The $f_H(C) \equiv 0$ means that the H$^+$(H$_3$O$^+$) is incapable of polarizing its neighboring HBs but only breaking and slightly repulsing its neighbors[37].
2) The $f_{Li}(C) \propto C$ means the constant shell size of the small Li$^+$ cation (radius = 0.78 Å) without being interfered with by other solutes. The constant slope indicates that the number of bonds per solute is conserved in the hydration shell. The electric field of a small Li$^+$ cation is fully screened by the H$_2$O dipoles in its hydration shells; thus, no cation–anion or cation–cation interaction is involved for the LiBr and HBr solutions.
3) The $f_{OH}(C) \propto C$ (<3100, 3610 cm$^{-1}$) means that the numbers of the O:⇔:O compression–elongated solvent H–O bonds ($f_{OH}(C) = 0.985C$) and the bond–order–deficiency shortened solute H–O bonds ($f_{OH}(C) = 0.322C$) are proportional to the solute concentration. Bond order deficiency shortens and stiffens the bonds between undercoordinated atoms [51].
4) The $f_{Br}(C) \propto 1-\exp(-C/C_0)$ toward saturation means the number of H$_2$O molecules in the hydration shells is insufficient to fully screen the Br$^-$ (radius = 1.96 Å) solute local electric field because of the geometric limitation to molecules packed in the crystal–like water. This number inadequacy may further evidence the well–ordered crystal–like solvent. The solute can thus interact with their alike –only anion–anion repulsion exists in the Br⁻ – based solutions to weaken the local electric field of Br⁻. Therefore, the $f_{Br}(C)$ increases approaching saturation, the hydration shells size turns to be smaller, which limits the solute capability of bond transition.

Therefore, the $f_x(C)$ and its slope give profound information not only on the solute–solute and solute–solvent interaction but also on the relative number of bonds transiting from the referential mode of water to the hydration, by ionic polarization or O:⇔:O compression.

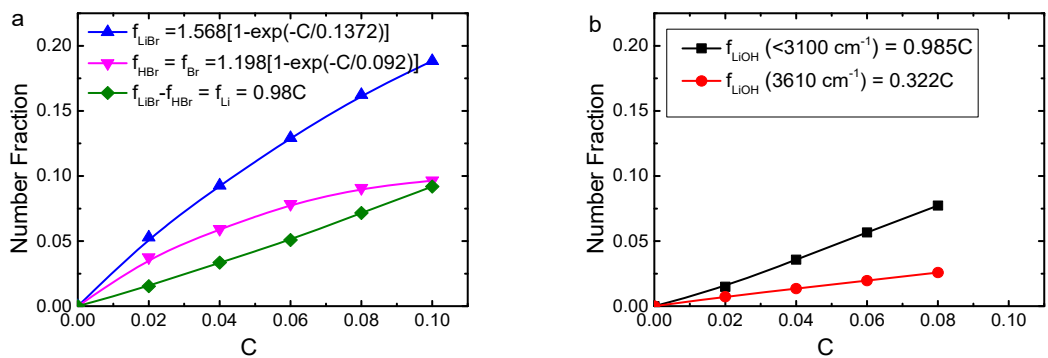

Figure 8. Concentration dependence of the fraction coefficients for LiBr/H₂O (3500 cm⁻¹), HBr/H₂O (3500 cm⁻¹), and LiOH (<3100, 3610 cm⁻¹) solutions. The liner $f_x(C)$ indicates the invariance of the Li⁺ and OH⁻ hydration shell size and the exponential $f_x(C)$ features Br⁻–water interaction with contribution of Br⁻ – Br⁻ interaction.

4. Hydrogen Bond Transition versus Solution Property

4.1. Surface Stress, Solution Viscosity, Molecular Diffusivity

Figure 9 a compares the concentration dependence of the contact angle between the solutions and glass substrate measured at 298 K. The surface stress is proportional to the contact angle. One can ignore the reaction between the glass surface and the solution, as we want to know the concentration trends of the stress change at the air–solution interface of a specific solution. Ionic polarization and O:⇔:O compression enhance the stress, but the H↔H point fragilization destructs the stress as the Li⁺ hydration forms an independent hydrating fragment. The H↔H fragmentation has the same effect of thermal fluctuation on depressing the surface stress[45] with different mechanisms. Thermal excitation weakens the individual O:H bond throughout the bulk water, but H↔H fragilization weakens the O:H bonds between the $Li^+$ hydrated fragments. Ionic and O:⇔:O polarization has the same effect of molecular undercoordination on constructing the surface stress. Both polarization and undercoordination form the supersolid phase; the former occurs in the hydration shell throughout the bulk, but the latter only takes place in skins. Ions may prefer occupying the skin of the solution, which is a different situation.

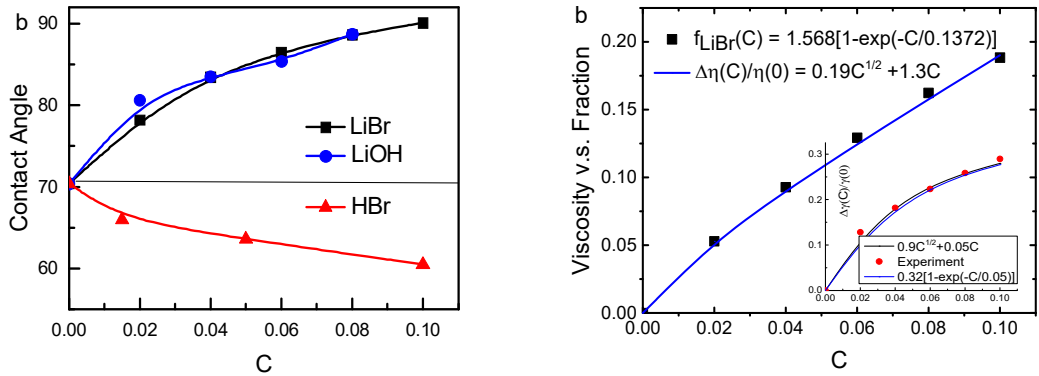

Figure 9. Concentration dependence of (a) solution contact angles on glass substrate and (b) trend agreement between the relative viscosity of Jones–Dole notion[91] and the $f_{LiBr}(C)$ contact angle for the LiBr/$H_2O$ solutions. Inset b shows that the LiBr/$H_2O$ surface stress follows the same exponential trend of the $f_{LiBr}(C)$ and the relative viscosity with different coefficients because of the additional molecular undercoordination effect that enhances the ionic polarization.

In aqueous solutions, solute molecules are taken as Brownian particles drifting randomly under thermal fluctuation by collision of the solvent molecules. The viscosity of salt solutions is one of the important macroscopic parameters often used to classify water–soluble salts into structure making or structure breaking. The drift motion diffusivity $D(\eta, R, T)$ and the solute–concentration–resolved solution viscosity $\eta(C)$ follow the Stokes–Einstein relation [92] and the Jones–Dole expression[91], respectively,

$$\begin{cases} \dfrac{D(\eta,R,T)}{D_0} = \dfrac{k_B T}{6\pi\eta R} & (Drift) \\ \dfrac{\Delta\eta(C)}{\eta(0)} = A\sqrt{C} + BC & (Viscosity) \end{cases}$$

where $\eta$, R, and $k_B$ are the viscosity, solute size, and Boltzmann constant, respectively. $D_0$ is the coefficient in pure water. The coefficient A and its nonlinear term is related to the solute mobility and solute–solute interaction. The coefficient B and the linear term reflects the solute–solvent molecular interactions. The $\eta(0)$ is the viscosity of neat water.

SFG measurements [93, 94] revealed that the $SCN^-$ and $CO_2$ solution viscosity increases with solute concentration or solution cooling. The H–O phonon relaxation time increases with the viscosity, and results in molecular motion dynamics. Therefore, ionic polarization stiffens the H–O phonon and slows down the molecular motion in the semirigid or supersolid structures.

One may note that the relative viscosity and the measured surface stress due to salt solvation are in the same manner of the $f_{LiBr}(C)$, $f_{LiBr}(C) = a[1-\exp(-C/C_0)]$. One can adjust the Jones–Dole viscosity coefficients A and B and fit the surface stress to match the measured $f_{LiBr}(C)$ curve in Figure 9 b. The trend consistency clarifies that the linear term corresponds to $Li^+$ hydration shell size and the nonlinear part to the resultant of $Br^-$–water and $Br^-$–$Br^-$ interactions. It is clear now that both the solution viscosity and the surface stress are proportional to the extent of polarization or to the sum of O:H–O bonds in the hydration shells. Therefore, polarization raises the surface stress, solution viscosity and rigidity, H–O phonon frequency, and H–O phonon lifetime but decreases the molecular drift mobility, consistently by shortening the H–O bond and lengthening the O:H nonbond.

4.2. LiOH Exothermal Solvation

According to chemical bond theory [95], energy stores in the chemical bonds and the energy emission or absorption proceeds by bond relaxation –the equilibrium atomic distance and binding energy change [82]. Bond dissociation and bond elongation release energy but bond formation and bond contraction absorb energy, leading to the exothermic and endothermic reaction. Molecular diffusive motion or structure fluctuation only dissipate energy with negligible energy absorption or energy emission.

LiOH solvation undergoes the exo– and endo–thermic reactions besides thermal dissipation by structural fluctuation and molecular diffusion and non–adiabatic calorimetric detection. The endothermic processes include ($Q_{a,i}$):

i) The hydrating H–O bond contraction by $Li^+$ polarization,
ii) Solute H–O contraction by bond–order–deficiency,
iii) Solvent H–O thermal contraction by temperature increases.

The exothermic processes include ($Q_{e,j}$):

i) LiOH dissolution into $Li^+$ and $OH^-$
ii) Solvent H–O elongation by O:⇔:O compression,
iii) O:H elongation by $Li^+$ polarization and thermal excitation.

The total energy should conserve:

$$\sum_3 Q_{e,i}(C) - \sum_3 Q_{a,j}(C) - \sum_2 Q_{dis,l}(C) = 0$$

These exo– and endo-thermic processes shall compensate each other to a certain extent during solvation. We neglect the energy dissipation and the thermodynamics of $Li^+$ solvation, as LiBr dissociation and ionic polarization derive no apparent temperature change. One can focus on the exothermic solvent H–O elongation by O:⇔:O compression and solute H–O contraction by bond–order–deficiency and the temperature change and thus estimate the energy emitted by the H–O elongation.

To seek for the correspondence between the solution temperature T(C) and the $f_{LiOH}(C)$ solutions, an

*in–situ* solution calorimetric detection was conducted using a regular thermometer to monitor the solution temperature in a glass beaker under the ambient temperature of 25 °C [39]. The solution was stirred using a magnetic bar rotating in the beaker in 5 Hz frequency. Figure 10 plots the LiOH solution temperature T(C) with comparison of the bond transition fraction coefficients f(C) for the elongated solvent H–O bonds and for the solute H–O bond contraction upon LiOH solvation.

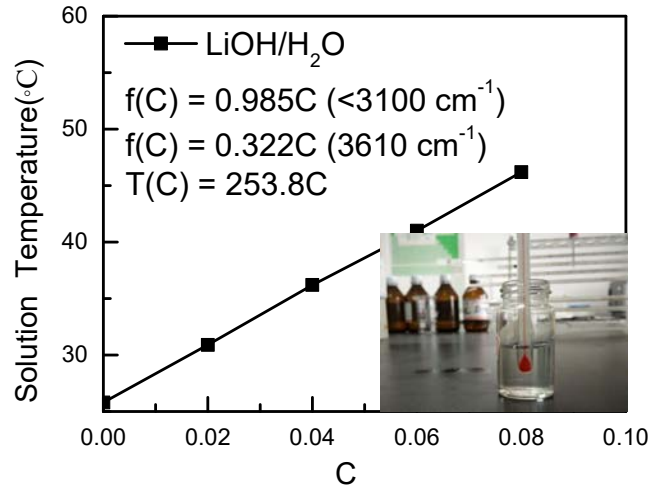

Figure 10. Linear correspondence between the highest solution temperatures T(C) and the concentration of the LiOH/H₂O solutions. Noted are the linear dependence of the fraction coefficient f(C) for the H–O phonon bands at <3100 cm$^{-1}$ and 3610 cm$^{-1}$ wave numbers (reprinted with permission from [39])

With the measured T(C) = 253.8C, $f_e$(C) = 0.985C (<3100 cm$^{-1}$), and $f_a$(C) = 0.322C (3610 cm$^{-1}$), one can estimate the energy emission from the solvent O:H–O bond elongated by O:⇔:O compression by ignoring the thermal fluctuation featured at 200 cm$^{-1}$. LiOH solvation transits the 200 cm$^{-1}$ mode for the O:H stretching (~0.095 eV[51]) partially to 110 and 300 cm$^{-1}$ which absorbs/emits a negligible amount of O:H energy.

The energy cost $Q_0$ to heat up a unit mass of the solution (m = 1) from $T_i$ to $T_f$ by increasing the solute contraction up to $C_M$ equals ($h_0$ = 4.18 J(gK)$^{-1}$ = 0.00039 eV(bondK)$^{-1}$ is the specific heat for liquid water):

$$\int_0^{Q_0} dq = h_0 \int_0^1 dm \int_0^{C_M} \frac{dt(C)}{dC} dC = h_0 T(C_M)$$

The energy difference between exothermic solvent H–O elongation $Q_e$ and the endothermic solute H–O contraction $Q_a$ heats up the solution ($h_e$ and $h_a$ are the energy emission and absorption per bond),

$$\int_0^{Q_e} dq_e - \int_0^{Q_a} dq_a = \left[ h_e \int_0^{f_e} dm_e - h_a \int_0^{f_a} dm_a \right] \int_{T_i}^{T_f} dt = [h_e f_e - h_a f_a] T(C_M)$$

Equaling the energy emission by H–O elongation to its loss by H–O contraction and heating solution with an approximation of $h_a \approx h_e$, yields,

$$h_e = \frac{h_0}{f_e - f_a h_a/h_e} \approx \frac{h_0}{(0.985 - 0.322)C_M} = 19 h_0$$

The energy emitted by an elongated H–O bond,

$$q_e = h_e T(C_M) = 253.8 C_M \times 19 h_0 = 0.151 (eV/bond),$$

which is 158% of the O:H cohesive energy 0.095 eV at room temperature.[51] It is thus verified that the energy remnant of the solvent H–O exothermic elongation and the solute H–O endothermic contraction heats up the solution.

We can also estimate the energy emission from the H–O bond elongation with the documented values of $(d_H, E_H, \omega_H) = (1.0\ \text{Å}, 4.0\ \text{eV}, 3200\ \text{cm}^{-1})$ for the bulk water [96], and for basic hydration (1.05 Å, $E_2$, 2500/3000 cm$^{-1}$) using the frequency function, $\omega^2 \propto E/d^2$:

$$\Delta E = 2E \left( \frac{\Delta d}{d} + \frac{\Delta \omega}{\omega} \right) = 8 \left( \frac{0.05}{1} - \frac{(700;\ 200)}{3200} \right) = -(1.35;\ 0.35)\ eV/bond$$

(3)

The H–O bond elongation losses its cohesive energy from 4.0 by 0.35~1.35 eV. Although the 0.15 eV energy emission may be underestimated, the O:⇔:O compression elongated H–O elongation is certainly the intrinsic and dominant source for heating up the LiBr solution.

5. Salt Solution Room-temperature Compression Icing

5.1. $P_C$ and $T_C$ for Phase Transition

It is usual for other materials that mechanical compression shortens all the interatomic bonds and stiffens their Raman phonons such as carbon allotropes [97]. However, for water and ice, compression shortens the O–O distance and the O:H length but lengthens the H–O bond, towards O:H–O segmental length symmetrization transiting into the $X^{th}$ phase at some 59 GPa pressure and at almost full temperature regime [2, 98-100]. The O:H and the H–O are identical in length at 0.10 nm but have different bond energies at the VII-X and VIII-X boundaries [77]. Compression of the phase X shortens both the H–O and the O:H very slightly and simultaneously [101]. Accordingly, compression stiffens the O:H phonon from below 400 cm$^{-1}$ to its above but softens the H–O phonon from above 3200 cm$^{-1}$ to its below [102-106], because of the O:H–O bond cooperativity and O–O repulsivity [51, 107].

Salt solvation modulates not only the solubility of dissolving biological molecules such as DNA and proteins but also the surface stress and solution viscosity and solubility [108-114]. Salt solvation varies the critical temperatures $T_C$ and the critical pressures $P_C$, for phase transition. The gelation time, $t_C$ for energy accumulation also varies with the type and concentration of the solutes for transiting the colloid from sol to gel [115-117]. The combination of salt solvation and mechanical compression makes the situation much more complicated [38]. Low-temperature and high-pressure Raman spectroscopy of ice transition from the phase VII to phase VIII and subsequently to the symmetrical $X^{th}$ phase revealed that NaCl or LiCl solvation raises the $P_C$ for transiting the solution ice from phase VII to phase X compared to pure ice [118, 119]. Solvation of the LiCl in the LiCl/H$_2$O number ratio of 1/50 and 1/6 requires 30 and 85 GPa more pressures with respect to that required for transiting pure water in to ice X at 60 GPa. Meanwhile, salt hydration not only stiffens the $\omega_H$ but also lengthens the O–O distance and shortens the $\omega_H$ phonon lifetime [120, 121].

Compression elevated $P_C$ for the VII-VIII and the VIII-X phase transition is attributed to the O:H compression with an association of H–O elongation through the intrinsic Coulomb repulsion between electron pairs on adjacent oxygen anions [2, 100]. Another opinion for this kind of phase transition is that the pressure degenerates the symmetrical double-well potentials into one located midway between oxygen ions, terminating the proton quantum transitional tunneling, which locates the proton in the fixed position [98]. The salt hydration elevated $P_C$ for the VII-X and VIII-X phase transition is explained on the base of two-component phase diagram for salted-water. The presence of salt hinders proton ordering and O:H–O segmental length symmetrization [118-120].

This section shows that heating has the same, but mechanical compression has an opposite effect of salt solvation on the O:H–O bond relaxation. However, heating enhances the thermal fluctuation and depresses the solution surface stress, but salvation does it contrastingly. Phase transition needs excessive energy to recover the salvation-elongated O:H–O bond by raising the $P_{C1}$ and $P_{C2}$ simultaneously and the $\Delta P_C$ changes with the solute type in the order of I > Br > Cl > F ~ 0. However, NaI concentration performs differently with solute type because of the solute-solute interaction comes into play at higher concentrations. The $P_{C1}$ grows faster than the $P_{C2}$. The $P_{C1}$ proceeds along the P-T path of the Liquid-VI boundary and the $P_{C2}$ along the VI-VII boundary. $P_{C1}$ and $P_{C2}$ then merge at the liquid-VI-VII triple-phase junction of 3.3 GPa and 350 K.

5.2. P-T-E Joint Effect on Bond Energy
5.2.1. Electrification and Compression

The ionic polarization stretches the $H_2O$ dipoles by elongating the O:H nonbond and softens its phonon and relaxes the H–O bond contrastingly because of the O:H–O bond cooperativity. Therefore, solute electrification reduces the molecular size and enlarges their separations with an association of strong polarization of the nonbonding electron lone pairs, which is responsible for the skin hydrophobicity, viscoelasticity, stress, and solubility, etc. [51].

Mechanical icing of salt solutions proceeds in two steps [38]. Firstly, salt solvation stores energy into the H–O bond by contraction and the amount of energy storage varies with the solute type and concentration. Secondly, mechanical compression then has to reverse the relaxed O:H–O bond to the right O:H and H–O energies of pure water for the phase transition [2]. Compression stores energy to the O:H nonbond and release energy from the initially contracted H–O by H–O elongation. Moving along the pressure path at 298 K in the phase diagram, one will go through phases of Liquid, ice VI, ice VII, and cross their boundaries, towards phase X at even higher pressures. For the deionized water, the Liquid-VI and the VI-VII phase transitions occur at 1.33→1.14 GPa and at 2.23→2.17 GPa, respectively, with the pressure sharp falls at the phase boundary for structural relaxation and O–O repulsion attenuation [38, 52, 119].

5.2.2.   $T_{xC}$ and $P_{Cx}$ versus Bond Energy

O:H–O bond relaxation and electron polarization dictate the detectable quantities such as phonon frequencies $\omega_x$, O1s binding energy shift, and the $T_{xC}$ and $P_{Cx}$ for phase transition (subscript x denotes different phase transition), as well as macroscopic properties such as hydrophobicity, toughness, slipperiness, viscoelasticity [51]. Quantities of immediate concern are the $\omega_x$ and the $T_{xC}$ [51]:

$$\begin{cases} \omega_x \propto \sqrt{E_x/\mu_x}/d_x \\ T_{xC} \propto \sum_{x=H,L} E_{xC} \end{cases}$$

(1)

The O:H–O bond cooperativity means that if one segment becomes shorter, it will be stiffer, and its characteristic phonon undergoes a blue shift; the other segment of the O:H–O bond relaxes contrastingly. The $\omega_x$ depends intrinsically on the reduced mass $\mu_x$, segmental length $d_x$ and energy $E_x$, and molecular coordination environment, which is the advantage of the DPS in direct estimating the segmental length and energy from observations.

The $P_{Cx}$ and the $T_C$ for a phase transition is correlated to the O:H–O bond energy $E_{xC}(X = L, H)$:

$$T_C \propto \sum_{L,H} E_{xC} = \begin{cases} \sum_{L,H} \left( E_{x0} - s_x \int_{P_0}^{P_{C0}} p \frac{dd_x}{dp} dp \right) & \text{(a, neat H}_2\text{O)} \\ \sum_{L,H} \left( E_x - s_x \int_{P_0}^{P_C} p \frac{dd_x}{dp} dp \right) & \text{(b, solution)} \end{cases}$$

(2)

$E_{H0} < E_H$ means that the H–O bond for pure water is weaker than it is in the salted water at the same pressure (ambient $P_0$ = 100 kPa ~ 0) [38, 46]. In contrast, $E_{L0} > E_L$. The integrals are energies stored into the bonds by compression. The summation over both segments of the O:H–O bond. When the pressure increases from $P_0$ to the $P_{C0}$ for neat water and to the $P_C$ for the salted, phase transition occurs. At transition, the bond energy equals to the difference between the two terms in the bracket. The change of the cross-section $s_x$ of the specific bond of $d_x$ length is assumed insignificant at relaxation with ignorance of the Poisson ratio. To raise the critical pressure from $P_{C0}$ to $P_C$ at the same critical temperature ($\Delta T_C = 0$ in the present situation), one needs excessive energy which is the difference between the bond energy of the salted and the neat water, therefore, eq (2) yields,

$$\Delta E_x - s_x \left( \int_{P_0}^{P_C} p \frac{dd_x}{dp} dp - \int_{P_0}^{P_{C0}} p \frac{dd_x}{dp} dp \right) = 0$$

where $\Delta E_x = E_x - E_{x0}$ is the energy stored into the O:H–O bond by salt solvation; compression from $P_{C0}$ to $P_C$ recovers the salt-induced distortion, and then phase transition occurs. This expression indicates that, if one wants to overcome the effect of solute electrification on the bond distortion for phase transition, one should increase the pressure from $P_{C0}$ to $P_C$ under the same critical temperature, say at 298 K, presently. The $\Delta E_x$ varies with the solute type and concentration in different manners, as the solute-solute interaction changes with solute concentration.

The segmental length $d_x$ varies with pressure in trends of [2],

$$\frac{dd_L}{dp}<0 \qquad \frac{dd_H}{dp}>0 \qquad \text{and } P_C > P_{C0},$$

which defines uniquely the segmental deformation energies derived by salt solvation,

$$\Delta E_x = s_x \left( \int_{P_0}^{P_C} p \frac{dd_x}{dp} dp - \int_{P_0}^{P_{C0}} p \frac{dd_x}{dp} dp \right) \begin{cases} >0 & (H-O) \\ <0 & (O:H) \end{cases}$$

(3)

Indeed, ionic polarization strengthens the H–O bond but weakens the O:H nonbond. The resultant of the $\Delta E_L$ loss and the $\Delta E_H$ gain governs the $\Delta P_C$ of the phase transition of the aqueous solutions.

5.3. Salting and Compressing

5.3.1. Hofmeister $P_{Cx}$ for NaX/H$_2$O Icing

High-pressure Raman examination of the NaX and concentrated Na/H$_2$O Liquid-VI and the VI-VII phase transition confirmed consistently the above predictions [38]. Figure 11 shows the typical H$_2$O and NaI/H$_2$O phonon cooperative relaxation dynamics at 298 K. Table 2 records the variation of the $P_{Cx}$ for each type of the solutions. The abrupt changes of the spectral shapes correspond to the $P_{Cx}$ for phase transition with structural relaxation. The $P_C$ sharp fall at phase boundary indicates that the structural relaxation weakens the O–O repulsion. Raman spectra for the room-temperature salt solution Liquid-VI and then VI-VII icing revealed the following [38]:

1) Three pressure zones from low to high correspond to phases of Liquid, ice VI, and ice VII, toward phase X [51].
2) Compression shortens the O:H nonbond and stiffens its $\omega_L$ phonon but does the opposite to the H–O bond over the full pressure range except for the $P_{Cx}$ at phase boundaries.
3) At transition, the gauged pressure drops, resulting from geometric restructuring that weakens the O–O repulsion; both the O:H and the H–O contract abruptly when cross the phase boundaries.

4) Most strikingly, the P$_{Cx}$ increases with the anion radius or the electronegativity difference between Na and X, following the Hofmeister series order: I > Br > Cl > F ≈ 0.

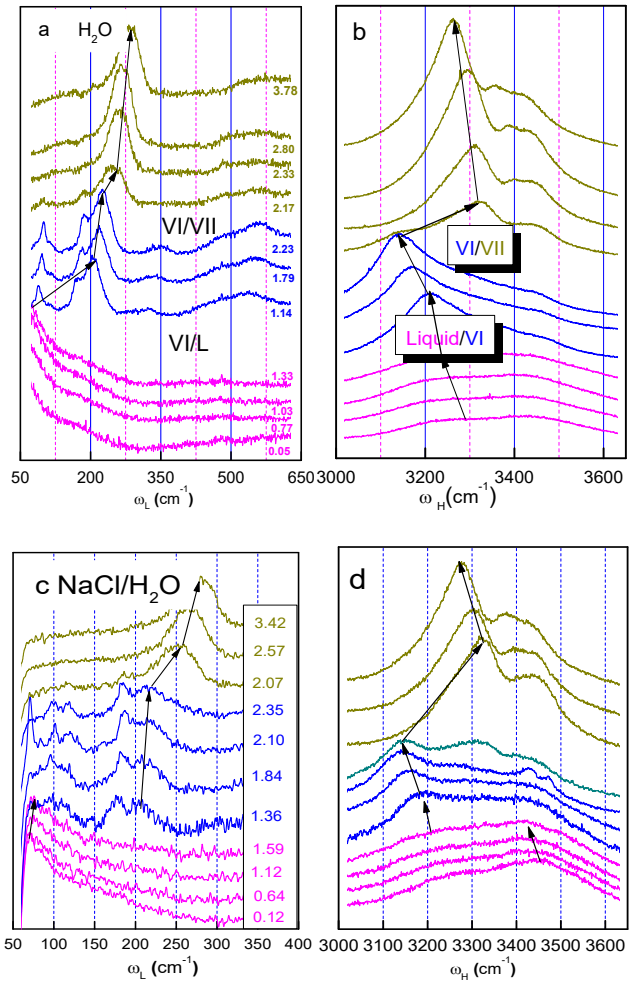

Figure 11. (a) O:H and H–O phonon cooperative relaxation of pure $H_2O$ under 298 K mechanical compression. Liquid-VI and VI-VII phase transition takes place at P$_C$ associated with fine shifts at the phase boundaries (reprinted with copyright permission from [38]).

Table 2. O:H–O bond relaxation dynamics of the NaX solutions transiting from Liquid to VI and VII phases under compression and ambient temperature. η is the elemental electronegativity. [a,b]

|  | $H_2O$ | NaF | NaCl | NaBr | NaI |
|---|---|---|---|---|---|
| Δη (η$_{Na}$ = 0.9) | - | 3.1 | 2.1 | 1.9 | 1.6 |
| R (R$_{Na+}$ = 0.98 Å) | - | 1.33 | 1.81 | 1.96 | 2.20 |

| | | | | | | |
|---|---|---|---|---|---|---|
| Liquid | $\Delta\omega_L$ | >0 | | | | |
| VI | | | | | | |
| VII | $\Delta\omega_H$ | <0 | | | | |
| L→VI boundary | $P_{C1}$ | 1.33→1.14 | 1.45→1.13 | 1.59→1.36 | 1.56→1.51 | 1.94→1.74 |
| | $\Delta\omega_L$ | >0 | | | | |
| | $\Delta\omega_H$ | <0 | | | | |
| VI→VII boundary | $P_{C2}$ | 2.23→2.17 | 2.22→2.06 | 2.35→2.07 | 2.79→2.71 | 3.27→2.98 |
| | $\Delta\omega_L$ | >0 | | | | |
| | $\Delta\omega_H$ | >0 | | | | |

[a] The critical pressures $P_{C1}$ and $P_{C2}$ vary with the solute type in the Hofmeister series order. An anion of larger radius and lower electronegativity raises more the $P_{Cx}$.

[b] The pressure abruption at transition indicates the weakening of the inter-oxygen Coulomb repulsion, shortening the O:H and the H–O spontaneously at phase boundaries.

### 5.3.2. NaI Concentration Resolved $P_{Cx}$

The concentration dependence of the $P_{Cx}$ for the Liquid-VI and VI-VII transition for the concentrated NaI/H$_2$O solutions at 298 K shows consistently that compression shortens the O:H nonbond and stiffens its phonons but the H–O bond responds to pressure contrastingly throughout the course unless at the phased boundaries [46]. At higher concentrations, say 0.05 and 0.10, the skin 3450 cm$^{-1}$ mode are more active in responding to pressure, which evidence the preferential skin occupancy of the I$^-$ anions that enhances the local electric filed.

The high-pressure Raman spectra from the concentrated NaI/H$_2$O solutions revealed the following [46]:

1) The $P_{C1}$ for the Liquid-VI transition increases faster than the $P_{C2}$ with NaI concentration increase till its maximum at 3.0 GPa and 0.10 concentration. $P_{C1}$ and $P_{C2}$ approach eventually to the triple phase junction at 3.3 GPa and 350 K.

2) The $P_{C2}$ for VI-VII transition changes insignificantly with concentration, remaining almost at the VI-VII boundary in the phase diagram, which contrasts with the trend of solution type.

3) The concentration trend of $P_{C1}$ along the L-VI boundary is equivalent to the simultaneous compressing and heating in the phase diagram.

The discrepancy between the solute type and the concentration on the critical pressures for phase transition arises from the involvement of anion-anion interaction that weakens the electric field at higher concentrations. Solute type determines the nature and the extent of the initial electrification; concentration increase modulates the local electric field and the extent of the initial H–O bond energy storage.

Furthermore, the $P_{C2}$ is less sensitive than the $P_{C1}$ to the change of solute concentration. One can imagine that the highly-compressed O:H–O bond is less sensitive to the local electric field of the hydration shells. They could be harder to deform further than those less-deformed under the same pressure.

Table 3. NaI/H$_2$O concentration dependence of the critical pressures for transiting from Liquid to the phase VI and then to the VII phase under compression (unit in GPa).

|  | 0 | 0.009 | 0.016 | 0.033 | 0.050 | 0.100 |
|---|---|---|---|---|---|---|
| H$_2$O/NaI | - | 111 | 62 | 30 | 20 | 10 |
| $P_{C1}$ | 1.33→1.14 | 1.82→1.72 | 1.94→1.74 | 2.21→2.03 | 2.39→2.29 | 3.05→2.89 |
| $P_{C2}$ | 2.23→2.17 | 2.90→2.61 | 3.27→2.98 | 2.88→2.87 | 2.90→2.61 | |

One may note that, the O:H–O bond is very sensitive to the environment such as pressure holding time, temperature, and phase precipitation, the measurement may not be readily reproducible, but the trends of measurements and the physical origin retain. The maximal $P_{C2}$, 3.27 > 3.05 > 2.23 GPa, for the 0.016 NaI/H$_2$O solution [38], the 0.10 NaI/H$_2$O, and the deionize water, demonstrates clearly the Hofmeister effect, anion-anion interaction, on the critical pressures for the room-temperature phase transition, as compared in Figure 12 and Table 3.

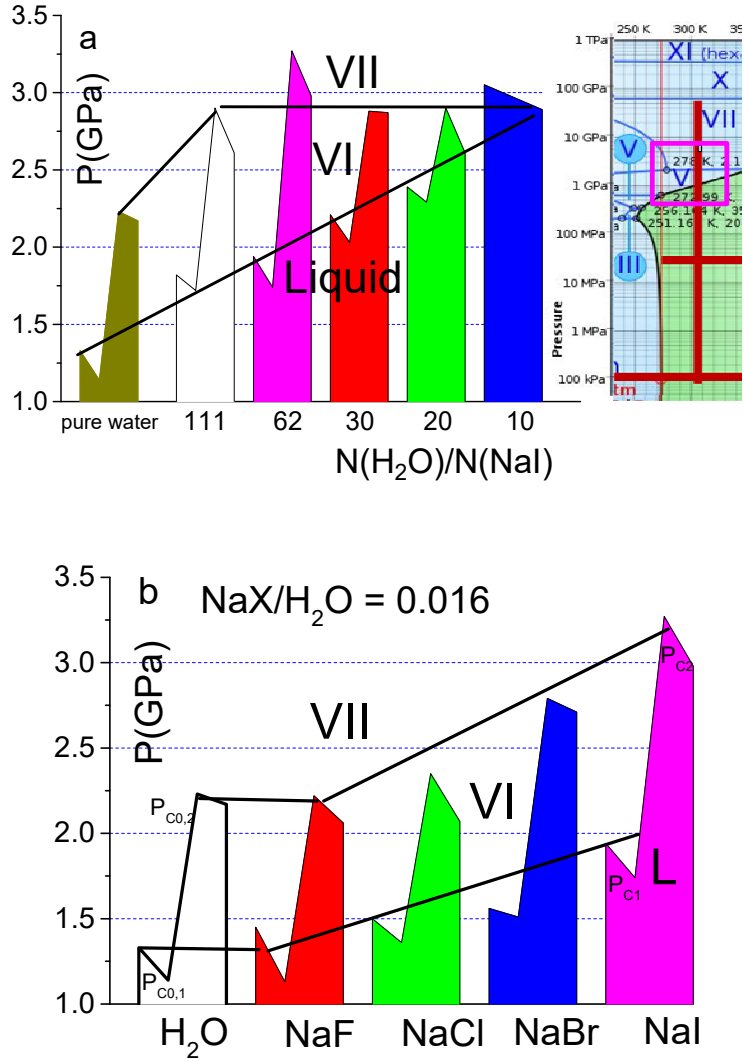

Figure 12. (a) NaI concentration and (b) NaX type [38] dependence of the critical $P_{C1}$ and $P_{C2}$ for the Liquid-VI and the VI-VII phase transition. Inset a shows the phase diagram with the pressure path at 298 K for the salt type. Indicated also the pressure path along the Liquid-VI boundary for the concentrated NaI solutions to the triple phase joint at 3.3 GPa and 350 K. Solute capabilities for raising the critical pressures at 0.016 molar ratio follow the Hofmeister order of I > Br > Cl > F = 0. The involvement of the anion-anion interaction discriminates the concentration from the type effect. The L-VI boundary is more sensitive than the VI-VII boundary to the solute concentration change.

5.3.3. Mechanical Compression and Ionic Polarization

Figure 13a shows that the $\omega_H$ phonon in ices of pure water and LiCl solution undergoes redshift under compression [120]. The extent of the H–O bond softening varies with its sites in the solution. The H–O stiffness relaxes less than it is in pure ice [122, 123] under compression, because of its initial contraction by ionic polarization [38, 46]. The initially shorter and stiffer H–O bonds under polarization are hardly compressible than the H–O bonds in pure water and ice.

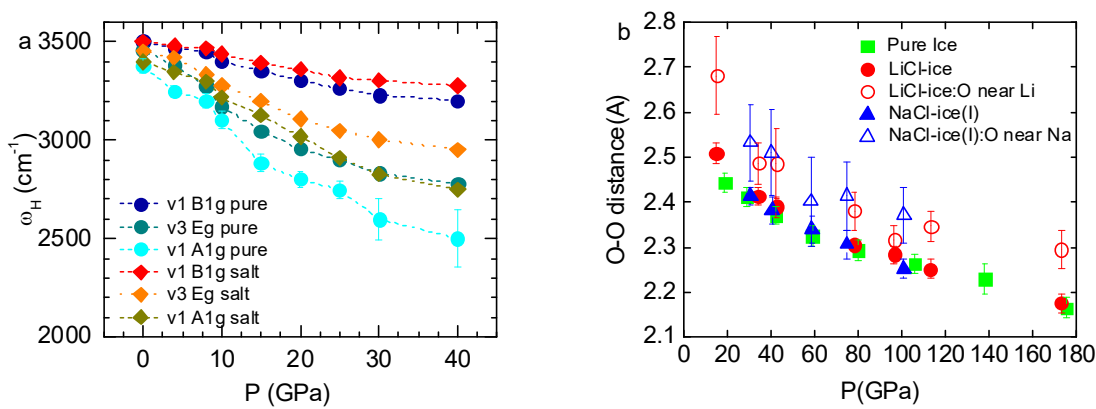

Figure 13. (a) Raman H–O stretching mode in pure and LiCl-contained ices (0.08 mol/kg [120]) and (b) the mean O–O distance in ices (squares), LiCl-contained ices (solid circles), and in NaCl ices (solid triangles). ν1 B1g > ν3 Eg > ν1 A1g mode. Open circles and triangles correspond to the mean O–O distances near the cation. (reprinted with permission from [121])

Figure 13b compares the mean O–O distance (O:H–O length) in pure ices, in LiCl- and NaCl-contained ices [121]. The O:H–O in the hydration shells of $Li^+$ and $Na^+$ is always longer than it is in the pure ice. The close situations for the pure ice, LiCl- and NaCl-contained ice suggest these signals arise from those O:H–O bonds outside the hydration shells. The O:H–O segmental length symmetrization could hardly happen even at 180 GPa for the O:H–O bonds with the hydration shells, in contrasting to the O:H–O symmetrization at 60 GPa pressure [2].

These observations evidence the compensation effect of mechanical compression and ionic polarization on the O:H–O bond cooperative relaxation. Compression shortens and stiffens the O:H nonbond while lengthens and softens the H–O bond, but the ionic polarization does it oppositely in salt solutions.

6. Phase transition of Confined Ice

Referring to Table 1 that features the multifield effect on the O:H–O segmental length and stiffness, Debye temperatures, quasisolid or quasiliquid (QS or QL) phase boundary (close to $T_m$ and $T_N$) dispersion. The H–O energy $E_H$ governs the $T_m$ and the O:H energy $E_L$ dictates the $T_N$. Compression and QS heating share the same effect, and molecular undercoordination and liquid heating share the contrasting effect on O:H–O bond relaxation and QS phase boundary dispersion. These regulations aid one to understand the confinement-compression modulated pressures and temperatures for the Solid-QS and QS-Liquid transition of water and ice.

Molecular undercoordination disperses the QS boundary outwardly, which raised the $T_M$ for a monolayer and the skin of bulk to 325 K [62] and 310 K [59] and lowers the $T_N$ for a 4.4, 3.4, 1.4 and 1.2 nm sized droplet to 242 K [124], 220 K [124], 205 K [55] and 172 K [64], respectively. For a cluster of 18 molecules or less, the $T_N$ is about 120 K [63]. Ionic polarization has an identical effect of molecular undercoordination on the O:H–O bond relaxation and the QS phase dispersion [4]. This fact explains why the solid-QS phase coexists between 299 and 333 K under compression, instead of 258 and 277 K for bulk water. However, compression disperses the QS boundary contrastingly, which raises the $T_N$ and lowers the $T_M$. Compression up to a pressure of 220 MPa lowers the $T_M$ for the liquid-QS transition of bulk water from 273 to 250 K, and this process is fully reversible, coined regelation [52].

Sotthewes and coworkers [61, 125] examined the multifield effect of mechanical compression, thermal excitation, and molecular undercoordination on the phase transition of ice with important discoveries. The atomic force microscopy showed in Figure 14a-c revealed the reversible transition from two-dimensional (2D) ice into a QS and Liquid phase confined between graphene and muscovite mica by compression. These dedicated observations are consistent with the prediction of QS boundary dispersivity by compression, QS heating, and molecular undercoordination, as illustrated in Figure 2c. Corresponding to the density extremes, or close to the $T_M$ for Liquid-QS and $T_N$ for Solid-QS phase transition, the QS boundaries are retractable by the Debye temperature and the characteristic vibration frequency shift, $\Theta_{Dx} \propto \omega_x$, cooperative relaxation [4].

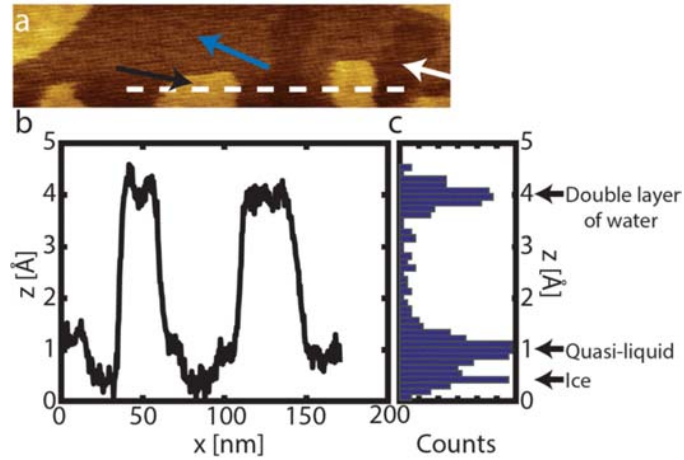

Figure 14. Height distribution of confined water under an external applied pressure higher than 6 GPa. (a) AFM topographic image (230 × 60 nm$^2$) of a melted fractal (~10 GPa). Three levels are present in the height profiles, namely, (i) the ice layer (denoted with a white arrow), (ii) the quasi-liquid layer (blue arrow), and (iii) the double water layer (black arrow). (b) Cross section across the white dashed line marked in panel (a). The height difference between the fractal and the double layer of water is about 3.6 Å. The quasi-liquid-like layer is ~70 ± 5 pm higher than the ice layer. (c) Histogram of the cross section in (b), showing three distinct peaks corresponding to the three different layers. (reprinted with copyright permission from [61] and [126])

Figure 15 shows two facts as consequence of compression-undercoordination heating on the $P_C$ and $T_N$ for solid-QS transition or homogeneous ice formation:

1. At room temperature, the critical pressure $P_C$ for the confined solid-QS transition amounts at 6 GPa that is much higher than the $P_C$ at 1.33 GPa for bulk Liquid-solid transition and the $P_C$ at 3.5 GPa for NaI solution of 0.1 molar concentration [38, 46].

2. The $P_C$ drops at the substrate heating. The ice and QS phase coexistence line appears at temperature between 293 and 333 K, which is higher than the QS phase between 258 and 277 K for bulk water at the ambient pressure [4].

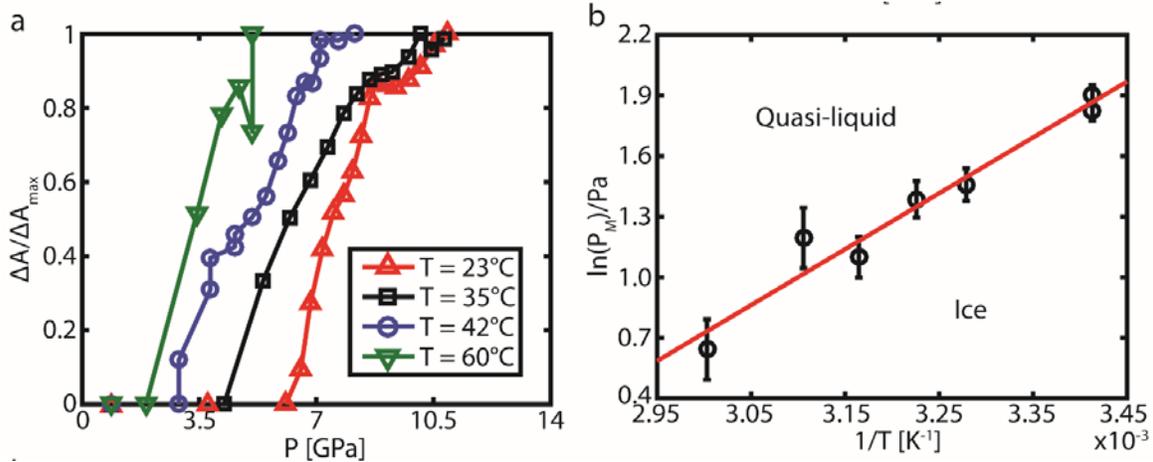

Figure 15. Compensation of substrate heating and mechanical compression on the (a) phase transition of confined water and the (b) $P_C$-T correlation. (reprinted with copyright permission from [61])

According to Table 1 regulations, one needs to shorten and stiffen the O:H nonbond by compression to raise the $T_N$ for translating the room-temperature confined ice into the QS phase at 299 K and below, as observed [61]. Compression up to 6 GPa raises the $T_N$ from far below to the room-temperature of confined ice. On the other hand, within the QS phase, the H–O bond follows the regular rule of thermal expansion - heating lengthens the H–O bond and shortens the O:H nonbond, which eases the O:H compression. Therefore, QS heating lowers the $P_C$ for the solid-QS transition. QS heating and compression compensate each other and thus they both soften the H-O bond to lower the $T_M$ and stiffen the O:H to raise the $T_N$.

7. Summary

We have thus resolved the multifield effect on the HB network and properties of water and Lewis-Hofmeister solutions in terms of $H^+(H_3O^+)$, $OH^-(:)$, $Y^+$, and $X^-$ injection. The DPS has allowed quantitative information on the number fraction and phonon stiffness transition from the mode of ordinary HBs to hydration. The multifield meditation of the O:H–O bonds in the network can thus be discriminated as follows:

1) $H_3O^+$ hydronium formation in acid solution creates the H↔H anti–HB that serves as a point breaker to disrupt the HBr solution network and the surface stress. The Br⁻ polarization dictates the O:H (from 200 partially to 110 and 300 cm$^{-1}$) and the H–O phonon (from 3200 to 3480 cm$^{-1}$) frequency cooperative shift. Acid solvation has the same effect of liquid heating on the O:H-O bond network and phonon relaxation through, respectively, H↔H fragilization and thermal fluctuation.

2) OH⁻ hydroxide forms the O:⇔:O super–HB point compressor to soften the nearest solvent H–O bond (from above 3100 cm$^{-1}$ to below), and meanwhile, the solute H–O bond shortens to its dangling radicals featured at 3610 cm$^{-1}$. The Y$^+$ polarization effect has been annihilated by the O:⇔:O compression and the solute H–O contraction. Base solvation has partly the effect of mechanical compression that lengthens and softens the H-O bond and shortens the O:H nonbond.

3) Y$^+$ and X⁻ ions serve each as a point polarizer that aligns, stretches, and polarizes the surrounding O:H–O bonds and makes the hydration shell supersolid. The polarization transits the $\omega_L$ from 200 to 100 cm$^{-1}$ and the $\omega_H$ from 3200 to 3480 cm$^{-1}$. Salt solvation has the same effect of molecular undercoordination to form the supersolid states.

4) The solute capability of bond transition follows: $f_H(C) = 0$, $f_Y(C) \propto f_{OH}(C) \propto C$, and $f_X(C) \propto 1-\exp(-C/C_0)$ toward saturation, which indicate the nonparaxiality of protons, invariant hydration volume size of Y$^+$ and HO⁻, and the variant X⁻ - X⁻ repulsion, respectively, and hence evidence the high structure order of $H_2O$ molecular in the solvent matrix.

5) The concentration trends consistent among the salt solution viscosity, surface stress, and the $f_{YX}(C)$ suggest their common origin of polarization associated with O:H–O bond transition from water to hydration shells.

6) The energy remnant of the solvent H–O exothermic elongation by O:⇔:O repulsion and the solute H–O endothermic contraction by bond–order–deficiency heats up the solution, which has little to do with the solute or solvent molecular motion dynamics.

7) Solvation and compression have the opposite effect on O:H–O relaxation and the shortened H–O bond by polarization can hardly elongated than the H–O bond in pure water. At constant concentration, the $T_C$ shows the Hofmeister sequences, but the involvement of the anion-anion repulsion alters the pattern of the critical pressure for the phase transition for the concentrated salt solutions.

8) Compression, confinement, and heating of the QS phase compensate one another other on the critical pressure of ice/QS transition at the ambient as compression always shortens the O:H nonbond while heating and confinement do oppositely in the QS phase of negative thermal expansion - H–O undergoes heating and compression elongation and O:H contraction but confinement does oppositely.

We have thus verified the essentiality of the O:⇔:O super–HB and the H↔H anti–HB dictating solute–solvent molecular interactions and their capabilities of transforming the HBs and the surface stress of these aqueous solutions systematically. The reported fine resolution and the consistent insight are only possible by taking the solutions as highly ordered, strongly correlated, and fluctuating systems with consideration of the O:H–O bond cooperativity. The solutes serve each as an impurity with their electric fields of polarization or repulsion and the local screening by the hydrating $H_2O$ dipoles. Observation may extend to general understanding of the solvation process of other acids, bases, and salts according to their abilities of creating the excessive $H^+$ protons, lone pairs, and ionic polarizers when they are hydrated. The nonbonding fragilization, compression, and polarization shall be critical to the hydration networks that functionalize DNA, proteins, cells, drugs, ionic channels, etc. The DPS forms such a powerful yet straightforward means to remove the artifacts and resolve the fraction and stiffness of the ordinary HBs transformation and their impact on solution properties.

**Acknowledgement**

Financial support from the Natural Science Foundation (Nos. 11872052; 21875024) and National Science Challenge Project (No. TZ2016001) of China are all gratefully acknowledged.

# References

[1] C.Q. Sun, Y. Sun, The Attribute of Water: Single Notion, Multiple Myths, Springer-Verlag, Heidelberg, 2016.
[2] C.Q. Sun, X. Zhang, W.T. Zheng, Chem Sci, 3 (2012) 1455-1460.
[3] C.Q. Sun, X. Zhang, J. Zhou, Y. Huang, Y. Zhou, W. Zheng, Journal of Physical Chemistry Letters, 4 (2013) 2565-2570.
[4] C.Q. Sun, X. Zhang, X. Fu, W. Zheng, J.-l. Kuo, Y. Zhou, Z. Shen, J. Zhou, Journal of Physical Chemistry Letters, 4 (2013) 3238-3244.
[5] C.Q. Sun, J. Chen, Y. Gong, X. Zhang, Y. Huang, Journal of Physical Chemistry B, 122 (2018) 1228-1238.
[6] Y.R. Shen, V. Ostroverkhov, Chemical reviews, 106 (2006) 1140-1154.
[7] H. Chen, W. Gan, B.-h. Wu, D. Wu, Y. Guo, H.-f. Wang, The Journal of Physical Chemistry B, 109 (2005) 8053-8063.
[8] M.E. Tuckerman, D. Marx, M. Parrinello, Nature, 417 (2002) 925-929.
[9] S.T. van der Post, C.S. Hsieh, M. Okuno, Y. Nagata, H.J. Bakker, M. Bonn, J. Hunger, Nat Commun, 6 (2015) 8384.
[10] A. Bragg, J. Verlet, A. Kammrath, O. Cheshnovsky, D. Neumark, Science, 306 (2004) 669-671.
[11] J. Verlet, A. Bragg, A. Kammrath, O. Cheshnovsky, D. Neumark, Science, 307 (2005) 93-96.
[12] S. Arrhenius, Nobel Lecture, (1903).
[13] J. Brönsted, Transactions of the Faraday Society, 24 (1928) 630-640.
[14] T.M. Lowry, I.J. Faulkner, Journal of the Chemical Society, Transactions, 127 (1925) 2883-2887.
[15] G.N. Lewis, Journal of the Franklin Institute, 226 (1938) 293-313.
[16] D. Chandler, Annual review of physical chemistry, 68 (2017) 19-38.
[17] F. Hofmeister, Arch. Exp. Pathol. Pharmacol, 24 (1888) 247-260.
[18] P. Jungwirth, P.S. Cremer, Nature chemistry, 6 (2014) 261-263.
[19] W.M. Cox, J.H. Wolfenden, Proc Roy Soc London A, 145 (1934) 475-488.
[20] P. Ball, J.E. Hallsworth, PCCP, 17 (2015) 8297-8305.
[21] K.D. Collins, M.W. Washabaugh, Quarterly Reviews of Biophysics, 18 (1985) 323-422.
[22] K.D. Collins, Biophysical Journal, 72 (1997) 65-76.
[23] K.D. Collins, Biophysical chemistry, 167 (2012) 43-59.
[24] H. Zhao, D. Huang, PLoS ONE, 6 (2011) e19923.
[25] X. Liu, H. Li, R. Li, D. Xie, J. Ni, L. Wu, http://www.nature.com/srep/2013/131021/srep03005/metrics, 4 (2014) 5047.
[26] W.B. O'Dell, D.C. Baker, S.E. McLain, PLoS ONE, 7 (2012) e45311.
[27] C. de Grotthuss, Galvanique. Ann. Chim, (1806).
[28] A.E. Stearn, H. Eyring, The Journal of Chemical Physics, 5 (1937) 113-124.
[29] G. Wannier, Annalen der Physik, 416 (1935) 545-568.
[30] M.L. Huggins, The Journal of Physical Chemistry, 40 (1936) 723-731.
[31] A. Hassanali, F. Giberti, J. Cuny, T.D. Kuhne, M. Parrinello, Proc Natl Acad Sci U S A, 110 (2013) 13723-13728.
[32] M. Eigen, Angewandte Chemie International Edition in English, 3 (1964) 1-19.
[33] G. Zundel, P. Schuster, G. Zundel, C. Sandorfy, Recent developments in theory and experiments, 2 (1976).
[34] D. Marx, M.E. Tuckerman, J. Hutter, M. Parrinello, Nature, 397 (1999) 601-604.
[35] C. Drechsel-Grau, D. Marx, Physical Chemistry Chemical Physics, 19 (2017) 2623-2635.
[36] J.A. Fournier, W.B. Carpenter, N.H.C. Lewis, A. Tokmakoff, Nature Chemistry, (2018).

[37] X. Zhang, Y. Zhou, Y. Gong, Y. Huang, C. Sun, Chemical Physics Letters, 678 (2017) 233-240.
[38] Q. Zeng, T. Yan, K. Wang, Y. Gong, Y. Zhou, Y. Huang, C.Q. Sun, B. Zou, Physical Chemistry Chemical Physics, 18 (2016) 14046-14054.
[39] C.Q. Sun, J. Chen, X. Liu, X. Zhang, Y. Huang, Chemical Physics Letters, 696 (2018) 139-143.
[40] Editorial, Science, 309 (2005) 78-102.
[41] Y. Zhou, Y. Huang, L. Li, Y. Gong, X. Liu, X. Zhang, C.Q. Sun, Vibrational Spectroscopy, 94 (2018) 31-36.
[42] Y. Zhou, D. Wu, Y. Gong, Z. Ma, Y. Huang, X. Zhang, C.Q. Sun, Journal of Molecular Liquids, 223 (2016) 1277-1283.
[43] Y. Zhou, Y. Huang, Z. Ma, Y. Gong, X. Zhang, Y. Sun, C.Q. Sun, Journal of Molecular Liquids, 221 (2016) 788-797.
[44] Y. Gong, Y. Zhou, H. Wu, D. Wu, Y. Huang, C.Q. Sun, Journal of Raman Spectroscopy, 47 (2016) 1351–1359.
[45] X. Zhang, T. Yan, Y. Huang, Z. Ma, X. Liu, B. Zou, C.Q. Sun, Physical Chemistry Chemical Physics, 16 (2014) 24666-24671.
[46] Q. Zeng, C. Yao, K. Wang, C.Q. Sun, B. Zou, PCCP, 19 (2017) 26645-26650
[47] M. Heyden, J. Sun, S. Funkner, G. Mathias, H. Forbert, M. Havenith, D. Marx, Proceedings of the National Academy of Sciences, 107 (2010) 12068-12073.
[48] J.A. Fournier, W. Carpenter, L. De Marco, A. Tokmakoff, Journal of the American Chemical Society, 138 (2016) 9634-9645.
[49] K. Tielrooij, S. Van Der Post, J. Hunger, M. Bonn, H. Bakker, The Journal of Physical Chemistry B, 115 (2011) 12638-12647.
[50] S. Funkner, G. Niehues, D.A. Schmidt, M. Heyden, G. Schwaab, K.M. Callahan, D.J. Tobias, M. Havenith, Journal of the American Chemical Society, 134 (2011) 1030-1035.
[51] Y.L. Huang, X. Zhang, Z.S. Ma, Y.C. Zhou, W.T. Zheng, J. Zhou, C.Q. Sun, Coordination Chemistry Reviews, 285 (2015) 109-165.
[52] X. Zhang, P. Sun, Y. Huang, T. Yan, Z. Ma, X. Liu, B. Zou, J. Zhou, W. Zheng, C.Q. Sun, Progress in Solid State Chemistry, 43 (2015) 71-81.
[53] S.A. Harich, X. Yang, D.W. Hwang, J.J. Lin, X. Yang, R.N. Dixon, Journal of Chemical Physics, 114 (2001) 7830-7837.
[54] S.A. Harich, D.W.H. Hwang, X. Yang, J.J. Lin, X. Yang, R.N. Dixon, The Journal of chemical physics, 113 (2000) 10073-10090.
[55] F. Mallamace, C. Branca, M. Broccio, C. Corsaro, C.Y. Mou, S.H. Chen, Proceedings of the National Academy of Sciences of the United States of America, 104 (2007) 18387-18391.
[56] X. Zhang, Y. Huang, P. Sun, X. Liu, Z. Ma, Y. Zhou, J. Zhou, W. Zheng, C.Q. Sun, Sci Rep, 5 (2015) 13655.
[57] J. Day, J. Beamish, Nature, 450 (2007) 853-856.
[58] C.Q. Sun, Progress in Solid State Chemistry, 35 (2007) 1-159.
[59] X. Zhang, Y. Huang, Z. Ma, Y. Zhou, W. Zheng, J. Zhou, C.Q. Sun, Physical Chemistry Chemical Physics, 16 (2014) 22987-22994.
[60] H. Sun, Journal of Physical Chemistry B, 102 (1998) 7338-7364.
[61] K. Sotthewes, P. Bampoulis, H.J. Zandvliet, D. Lohse, B. Poelsema, ACS nano, 11 (2017) 12723-12731.
[62] H. Qiu, W. Guo, Physical Review Letters, 110 (2013) 195701.
[63] R. Moro, R. Rabinovitch, C. Xia, V.V. Kresin, Physical Review Letters, 97 (2006) 123401.
[64] F.G. Alabarse, J. Haines, O. Cambon, C. Levelut, D. Bourgogne, A. Haidoux, D. Granier, B. Coasne, Physical Review Letters, 109 (2012) 035701.
[65] B. Wang, W. Jiang, Y. Gao, B.K. Teo, Z. Wang, Nano Research, 9 (2016) 2782-2795.
[66] H. Bhatt, A.K. Mishra, C. Murli, A.K. Verma, N. Garg, M.N. Deo, S.M. Sharma, Physical Chemistry Chemical

Physics, 18 (2016) 8065-8074.

[67] H. Bhatt, C. Murli, A.K. Mishra, A.K. Verma, N. Garg, M.N. Deo, R. Chitra, S.M. Sharma, The Journal of Physical Chemistry B, 120 (2016) 851-859.

[68] D. Kang, J. Dai, H. Sun, Y. Hou, J. Yuan, http://www.nature.com/srep/2013/131021/srep03005/metrics, 3 (2013).

[69] K. Dong, S. Zhang, Q. Wang, Science China Chemistry, 58 (2015) 495-500.

[70] F. Li, Z. Men, S. Li, S. Wang, Z. Li, C. Sun, Spectrochimica Acta Part A: Molecular and Biomolecular Spectroscopy, 189 (2018) 621-624.

[71] F.B. Li, Z.L. Li, S.H. Wang, S. Li, Z.W. Men, S.L. Ouyang, C.L. Sun, Spectrochimica Acta Part a-Molecular and Biomolecular Spectroscopy, 183 (2017) 425-430.

[72] X. Zhang, Y. Huang, Z. Ma, Y. Zhou, J. Zhou, W. Zheng, Q. Jiang, C.Q. Sun, Physical Chemistry Chemical Physics, 16 (2014) 22995-23002.

[73] T.F. Kahan, J.P. Reid, D.J. Donaldson, Journal of Physical Chemistry A, 111 (2007) 11006-11012.

[74] M.X. Gu, C.Q. Sun, Z. Chen, T.C.A. Yeung, S. Li, C.M. Tan, V. Nosik, Physical Review B, 75 (2007) 125403.

[75] F. Perakis, K. Amann-Winkel, F. Lehmkühler, M. Sprung, D. Mariedahl, J.A. Sellberg, H. Pathak, A. Späh, F. Cavalca, D. Schlesinger, A. Ricci, A. Jain, B. Massani, F. Aubree, C.J. Benmore, T. Loerting, G. Grübel, L.G.M. Pettersson, A. Nilsson, Proceedings of the National Academy of Sciences, 114 (2017) 8193-8198.

[76] J.A. Sellberg, C. Huang, T.A. McQueen, N.D. Loh, H. Laksmono, D. Schlesinger, R.G. Sierra, D. Nordlund, C.Y. Hampton, D. Starodub, D.P. DePonte, M. Beye, C. Chen, A.V. Martin, A. Barty, K.T. Wikfeldt, T.M. Weiss, C. Caronna, J. Feldkamp, L.B. Skinner, M.M. Seibert, M. Messerschmidt, G.J. Williams, S. Boutet, L.G. Pettersson, M.J. Bogan, A. Nilsson, Nature, 510 (2014) 381-384.

[77] Y. Huang, X. Zhang, Z. Ma, Y. Zhou, G. Zhou, C.Q. Sun, Journal of Physical Chemistry B, 117 (2013) 13639-13645.

[78] Y.L. Huang, X. Zhang, Z.S. Ma, G.H. Zhou, Y.Y. Gong, C.Q. Sun, Journal of Physical Chemistry C, 119 (2015) 16962-16971.

[79] J. Chen, C. Yao, X. Liu, X. Zhang, C.Q. Sun, Y. Huang, Chemistry Select, 2 (2017) 8517-8523.

[80] J. Sun, G. Niehues, H. Forbert, D. Decka, G. Schwaab, D. Marx, M. Havenith, Journal of the American Chemical Society, 136 (2014) 5031-5038.

[81] P. Cotterill, Progress in Materials Science, 9 (1961) 205-301.

[82] G.W. Shim, K. Yoo, S.-B. Seo, J. Shin, D.Y. Jung, I.-S. Kang, C.W. Ahn, B.J. Cho, S.-Y. Choi, ACS Nano, 8 (2014) 6655-6662.

[83] Y. Zhou, Y. Gong, Y. Huang, Z. Ma, X. Zhang, C.Q. Sun, Journal of Molecular Liquids, 244 (2017) 415-421.

[84] B. Wang, W. Jiang, Y. Gao, Z. Zhang, C. Sun, F. Liu, Z. Wang, RSC Advances, 7 (2017) 11680-11683.

[85] C.Q. Sun, in, United States, 2017.

[86] X.J. Liu, M.L. Bo, X. Zhang, L. Li, Y.G. Nie, H. TIan, Y. Sun, S. Xu, Y. Wang, W. Zheng, C.Q. Sun, Chemical reviews, 115 (2015) 6746-6810.

[87] Y. Zhou, Yuan Zhong, X. Liu, Y. Huang, X. Zhang, C.Q. Sun, Journal of Molecular Liquids, 248 (2017).

[88] X. Zhang, X. Liu, Y. Zhong, Z. Zhou, Y. Huang, C.Q. Sun, Langmuir, 32 (2016) 11321-11327.

[89] M. Thämer, L. De Marco, K. Ramasesha, A. Mandal, A. Tokmakoff, Science, 350 (2015) 78-82.

[90] A. Mandal, K. Ramasesha, L. De Marco, A. Tokmakoff, Journal of Chemical Physics, 140 (2014) 204508.

[91] G. Jones, M. Dole, Journal of the American Chemical Society, 51 (1929) 2950-2964.

[92] J.C. Araque, S.K. Yadav, M. Shadeck, M. Maroncelli, C.J. Margulis, The Journal of Physical Chemistry B, 119 (2015) 7015-7029.

[93] T. Brinzer, E.J. Berquist, Z. Ren, 任哲, S. Dutta, C.A. Johnson, C.S. Krisher, D.S. Lambrecht, S. Garrett-Roe, The Journal of chemical physics, 142 (2015) 212425.

[94] Z. Ren, A.S. Ivanova, D. Couchot-Vore, S. Garrett-Roe, The journal of physical chemistry letters, 5 (2014) 1541-1546.
[95] L. Pauling, The Nature of the Chemical Bond, 3 ed., Cornell University press, Ithaca, NY, 1960.
[96] Y. Huang, X. Zhang, Z. Ma, Y. Zhou, J. Zhou, W. Zheng, C.Q. Sun, Sci Rep, 3 (2013) 3005.
[97] W.-T. Zheng, C.Q. Sun, Energy & Environmental Science, 4 (2011) 627-655.
[98] M. Benoit, D. Marx, M. Parrinello, Nature, 392 (1998) 258-261.
[99] D. Kang, J. Dai, Y. Hou, J. Yuan, Journal of Chemical Physics, 133 (2010) 014302.
[100] S. Chen, Z. Xu, J. Li, New Journal of Physics, 18 (2016) 023052.
[101] X. Zhang, S. Chen, J. Li, Sci Rep, 6 (2016) 37161.
[102] Y. Yoshimura, S.T. Stewart, M. Somayazulu, H. Mao, R.J. Hemley, Journal of Chemical Physics, 124 (2006) 024502.
[103] P. Pruzan, J.C. Chervin, E. Wolanin, B. Canny, M. Gauthier, M. Hanfland, Journal of Raman Spectroscopy, 34 (2003) 591-610.
[104] M. Song, H. Yamawaki, H. Fujihisa, M. Sakashita, K. Aoki, Physical Review B, 60 (1999) 12644.
[105] Y. Yoshimura, S.T. Stewart, M. Somayazulu, H.K. Mao, R.J. Hemley, Journal of Physical Chemistry B, 115 (2011) 3756-3760.
[106] Y. Yoshimura, S.T. Stewart, H.K. Mao, R.J. Hemley, Journal of Chemical Physics, 126 (2007) 174505.
[107] N. Mishchuk, V. Goncharuk, Journal of Water Chemistry and Technology, 39 (2017) 125-131.
[108] L.M. Levering, M.R. Sierra-Hernández, H.C. Allen, Journal of Physical Chemistry C, 111 (2007) 8814-8826.
[109] L.M. Pegram, M.T. Record, The Journal of Physical Chemistry B, 111 (2007) 5411-5417.
[110] R.K. Ameta, M. Singh, Journal of Molecular Liquids, 203 (2015) 29-38.
[111] P. Lo Nostro, B.W. Ninham, Chemical reviews, 112 (2012) 2286-2322.
[112] C.M. Johnson, S. Baldelli, Chemical reviews, 114 (2014) 8416-8446.
[113] X.P. Li, K. Huang, J.Y. Lin, Y.Z. Xu, H.Z. Liu, Prog. Chem., 26 (2014) 1285-1291.
[114] E.K. Wilson, C&EN Archives., 90 (2012) 42-43.
[115] Y.R. Xu, L. Li, P.J. Zheng, Y.C. Lam, X. Hu, Langmuir, 20 (2004) 6134-6138.
[116] M. van der Linden, B.O. Conchúir, E. Spigone, A. Niranjan, A. Zaccone, P. Cicuta, The Journal of Physical Chemistry Letters, (2015) 2881-2887.
[117] F. Aliotta, M. Pochylski, R. Ponterio, F. Saija, G. Salvato, C. Vasi, Physical Review B, 86 (2012) 134301.
[118] G.N. Ruiz, L.E. Bove, H.R. Corti, T. Loerting, Physical Chemistry Chemical Physics, 16 (2014) 18553-18562.
[119] S. Klotz, L.E. Bove, T. Strässle, T.C. Hansen, A.M. Saitta, Nature Materials, 8 (2009) 405-409.
[120] L.E. Bove, R. Gaal, Z. Raza, A.A. Ludl, S. Klotz, A.M. Saitta, A.F. Goncharov, P. Gillet, Proc Natl Acad Sci U S A, 112 (2015) 8216–8220.
[121] Y. Bronstein, P. Depondt, L.E. Bove, R. Gaal, A.M. Saitta, F. Finocchi, Physical Review B, 93 (2016) 024104.
[122] A.F. Goncharov, V.V. Struzhkin, M.S. Somayazulu, R.J. Hemley, H.K. Mao, Science, 273 (1996) 218-220.
[123] A.F. Goncharov, V.V. Struzhkin, H.-k. Mao, R.J. Hemley, Physical Review Letters, 83 (1999) 1998.
[124] M. Erko, D. Wallacher, A. Hoell, T. Hauss, I. Zizak, O. Paris, PCCP, 14 (2012) 3852-3858.
[125] P. Bampoulis, K. Sotthewes, E. Dollekamp, B. Poelsema, Surface Science Reports, (2018).
[126] X. Zhang, P. Sun, Y. Huang, Z. Ma, X. Liu, J. Zhou, W. Zheng, C.Q. Sun, Journal of Physical Chemistry B, 119 (2015) 5265-5269.